\documentclass[aps,amsmath,amssymb,showpacs]{revtex4}
\usepackage{graphicx,bm,dcolumn}

\def\be{\begin{equation}}
\def\ee{\end{equation}}
\def\beqn{\begin{eqnarray}}
\def\eeqn{\end{eqnarray}}
\def\barr{\begin{array}}
\def\earr{\end{array}}
\def\btab{\begin{tabular}}
\def\etab{\end{tabular}}
\def\bite{\begin{itemize}}
\def\eite{\end{itemize}}
\def\bcen{\begin{center}}
\def\ecen{\end{center}}

\def\qt0{\tilde{q}_0}

\def\be{\begin{equation}}
\def\ee{\end{equation}}
\def\hate{\hat{\mathbf{e}}}
\def\bq{\mathbf{q}}
\def\bk{\mathbf{k}}
\def\bvare{\boldsymbol{\varepsilon}}
\def\bsig{\boldsymbol{\sigma}}

\def\sc{\scriptsize}

\begin{document}

\baselineskip=14pt

\input epsf.tex    

\input psfig.sty


\title{The Gerasimov-Drell-Hearn Sum Rule and the Spin Structure of the Nucleon}

\markboth{Drechsel \& Tiator}{The GDH Sum Rule and Nucleon Spin Structure}

\author{Dieter Drechsel and Lothar Tiator}
\affiliation{Institut f\"ur Kernphysik, Universit\"at Mainz, 55099 Mainz,
Germany}



\begin{abstract}
The Gerasimov-Drell-Hearn sum rule is one of several dispersive sum rules that
connect the Compton scattering amplitudes to the inclusive photoproduction
cross sections of the target under investigation. Being based on such universal
principles as causality, unitarity, and gauge invariance, these sum rules
provide a unique testing ground to study the internal degrees of freedom that
hold the system together. The present article reviews these sum rules for the
spin-dependent cross sections of the nucleon by presenting an overview of
recent experiments and theoretical approaches. The generalization from real to
virtual photons provides a microscope of variable resolution: At small
virtuality of the photon, the data sample information about the long range
phenomena, which are described by effective degrees of freedom (Goldstone
bosons and collective resonances), whereas the primary degrees of freedom
(quarks and gluons) become visible at the larger virtualities. Through a rich
body of new data and several theoretical developments, a unified picture of
virtual Compton scattering emerges, which ranges from coherent to incoherent
processes, and from the generalized spin polarizabilities on the low-energy
side to higher twist effects in deep inelastic lepton scattering.
\end{abstract}

\maketitle

\section*{1.\quad INTRODUCTION}

The Gerasimov-Drell-Hearn (GDH) sum rule relates the anomalous magnetic moment
(a.m.m.) of a particle to an energy-weighted integral over its photoabsorption
cross sections~\cite{Ger65}. This relation bears out that a finite value of the
a.m.m. requires the existence of an excitation spectrum, and that both
phenomena are simply different aspects of a particle with intrinsic degrees of
freedom. A further property of composite objects is their spatial extension in
terms of size and shape, which reveals itself through form factors measured by
elastic lepton scattering. The less known Burkhardt-Cottingham (BC) sum rule
connects a particular combination of these form factors to another
energy-weighted integral over the total absorption cross sections for virtual
photons~\cite{BC70}. The discovery of the proton's large a.m.m.~\cite{Ste33}
marked the beginning of hadronic physics. Today, 70 years later, we are still
struggling to describe the structure of strongly interacting particles in a
quantitative way.

Although quantum chromodynamics (QCD) is generally accepted as the underlying
theory of the strong interactions, its low-energy aspects are difficult to deal
with because of the large coupling constant in that region. However, the
low-energy region is the natural habitat of protons and neutrons, and therefore
of practically all visible matter in the universe. A plethora of models have
been inspired by QCD, but none can be quantitatively derived from it. Only two
descriptions are, in principle, exact realizations of QCD, namely chiral
perturbation theory (ChPT) and lattice gauge theory. The former is an expansion
in small external momenta and small (current) quark masses, and therefore it is
limited to the threshold region and to the light $u$, $d$, and $s$ quarks. The
effective degrees of freedom are the Goldstone bosons (quark-antiquark
oscillations), notably the pions; all other possible structures are absorbed in
contact interactions with ``low-energy constants'' (LECs) as proportionality
factors. Lattice gauge theory, on the other hand, prescribes how to evaluate
the QCD Lagrangian numerically on a space-time lattice in terms of quarks and
gluons. For computational reasons the calculations have been mostly performed
in the so-called quenched approximation, with only the minimal configuration of
three quarks in configuration space. With algorithms such as ``improved
actions'' for the light quarks and an increase in computing power by tera-flop
machines, it is slowly becoming possible to include sea-quark degrees of
freedom as well. Such effects have proved to be indispensable for an accurate
description of hadronic observables~\cite{Dav04}.

The persistent difficulties in quantitatively understanding nucleon structure
explain why we are interested in a relation like the GDH sum rule. Being based
on such general principles as causality, unitarity, and Lorentz and gauge
invariance, this sum rule should be valid for every model or theory respecting
these principles and having a ``reasonable'' high-energy behavior. Therefore,
the agreement or disagreement between theoretical predictions and the sum rule
provides invaluable information on the quality of the approximations involved
and whether or not the relevant degrees of freedom have been included.
Similarly, if we compare the sum rule value with accurate experimental data for
the photoabsorption cross sections up to a certain maximum energy, we learn
whether the physics responsible for the a.m.m. is also the physics of
photoproduction to that energy, or whether phenomena at higher energies and
possibly new degrees of freedom come into the game.

The a.m.m. $\kappa$ can be read off the relation between the total magnetic
moment $\vec{\mu}$ and the spin $\vec{S}$ of a particle in its center-of-mass
(c.m.) frame,
\be \label{mu} \vec{\mu} = \frac{e}{M}\,(Q+\kappa)\,\vec{S}\ , \ee
where $eQ$ is the charge and $M$ the mass of the particle. The elementary charge $e=|e|$
is defined by $e^2/4\pi = \alpha_{fs}\approx 1/137$. We use ``natural'' units,
$c=\hbar=1$, and the nuclear magneton is given by the ``normal'' magnetic moment of the
proton, $\mu_N=e/2M_p$. We also recall that the ratio between the magnetic moment
$\vec{\mu}$ and the orbital angular momentum $\vec{L}$ of a uniformly charged rotating
body is $eQ/2M$, while the ratio between the ``normal'' magnetic moment $\vec{\mu}$ and
the spin $\vec{S}$ takes twice that value. This factor is described by the gyromagnetic
ratio $g=2$ as predicted from Dirac's equation for a spin $1/2$ particle. According to
``Belinfante's conjecture''~\cite{Bel53} a particle of general spin $S$ should have
$g=1/S$. However, Ferrara et al.~\cite{Fer92} have quite convincingly shown that $g=2$ is
required for particles of any spin $S$ in order to obtain a reasonable high-energy
behavior.

The GDH sum rule relates the a.m.m. to the inclusive cross sections $\sigma_P$ and
$\sigma_A$ for spherically polarized photons impinging on particles with spin parallel
(P) or antiparallel (A) to the spin or helicity of the photon,
\be \label{e2kappa2} \frac{\pi e^2\kappa^2}{M^2}\,S = \int^{\infty}_{\nu_0}
\frac{d\nu}{\nu}\, \left(\sigma_P(\nu) - \sigma_A(\nu)\right )\ , \ee
where the integral runs over the photon energy in the laboratory frame, $\nu$,
from the lowest threshold $\nu_o$ to infinity. Because the left-hand side (LHS)
of this equation is positive, we can draw the qualitative conclusion that the
photon prefers to be absorbed with its spin parallel to the target spin. Of
course, the sum rule is much more quantitative. The same agent that is
responsible for the a.m.m. on the LHS of Eq.~(\ref{e2kappa2}) must also lead to
an appropriate energy dependence of the helicity difference of the cross
sections, $\sigma_P-\sigma_A$, such that the sum rule is fulfilled. However,
keep in mind one caveat: A basic assumption in deriving Eq.~(\ref{e2kappa2}) is
that $\sigma_P-\sigma_A$ vanishes sufficiently fast such that the integral
converges.

Having followed through to this point, the reader may well ask whether we can
ever be sure that the GDH sum rule is fulfilled. In a sense we cannot, because
ultimately physical questions can only be answered by experiment, and no
experiment will ever decide whether an infinite integral converges or diverges
in the mathematical sense. However, we should view such questions from a more
physical standpoint. First, the GDH sum rule is an ideal testing ground,
because the LHS of Eq.~(\ref{e2kappa2}) is given by ground state properties
and, in the case of the nucleon, is known to at least eight decimal places. The
result is $205~\mu$b for the proton and $233~\mu$b for the neutron. Thirty
years ago, the first estimates for the right-hand side (RHS) of
Eq.~(\ref{e2kappa2}) were $261~\mu$b for the proton and $183~\mu$b for the
neutron~\cite{Kar73}. Over the following years the predictions moved even
further away from the sum rule values despite an improving data basis, simply
because these data were not sensitive to the helicity difference of the
inclusive cross sections. Many explanations for the apparent violation of the
sum rule followed, but in view of the more recent experimental evidence we have
no intention to deliberate about these models.

Instead we recall the arguments of Ferrara et al.~\cite{Fer92} that the
gyromagnetic ratio $g$ takes its ``natural'' value of 2 for particles of any
spin, and that $g=2$ (at tree level) is the only value that allows for a
reliable perturbative expansion. The large a.m.m. of the nucleon does not
invalidate these arguments, because it is due to the composite structure of the
proton and neutron, which expresses itself by -- among other effects -- large
unitarity corrections from pion loops. Such spatially extended phenomena,
however, should not affect the high-energy limit of Compton scattering. Indeed,
it was observed long ago~\cite{Wei70} that the requirement of a well-behaved
high-energy scattering amplitude implies $g=2$ and points to the existence of
an unsubtracted dispersion relation. As far as composite systems of particles
are concerned, Brodsky \& Primack~\cite{Bro68} showed that the spin-dependent
current associated with the c.m. motion provides the additional interaction
terms to verify the GDH sum rule for bound states of any spin.

To further illustrate this point let us return to the small a.m.m. of the
electron, which can be evaluated in quantum electrodynamics (QED) to 10 decimal
places. Since the associated photon-electron loops are spread over a spatial
volume of $\sim10^6$~fm$^3$, a photon of a few hundred MeV or a wavelength
$\lambda\lesssim1$~fm will decouple from such a large object. Therefore the
a.m.m. should not affect the high-energy limit at all and, along the same
lines, the a.m.m. of the electron does not preclude its use as an ideal point
particle to study the form factors of the nucleon~\cite{Berm91}. A footnote in
the cited paper of Ferrara et al.~\cite{Fer92} makes room for another
interesting idea. These authors noticed that in a special class of
supersymmetric QED, the electron has $g=2$ even after radiative corrections. In
such a completely supersymmetric world, therefore, all GDH integrals would
vanish and all particles would be ``truly elementary'' and pointlike objects.
The finite value of the a.m.m. in the real world should then be interpreted as
a measure of supersymmetry-breaking effects.

For the reasons mentioned above, we cannot yet calculate the photoabsorption
cross section for hadronic systems from first principles. However, such
calculations have been performed in QED. The lowest order contribution to the
a.m.m. of the electron is the Schwinger correction $\kappa_e =
-\alpha_{fs}/(2\pi)$, and altogether the LHS of Eq.~(\ref{e2kappa2}) is of
order $e^6$ or $\alpha_{fs}^3$. Therefore, the RHS must vanish at order
$\alpha_{fs}^2$, which corresponds to Compton scattering at tree level. The
relevant finite terms of order $\alpha_{fs}^3$ are then provided by radiative
corrections to Compton scattering, pair production, and photon scattering with
the production of a second photon. When the RHS of Eq.~(\ref{e2kappa2}) was
evaluated by numerical integration and compared to the LHS, it was indeed
recently shown that the GDH sum rule holds in QED within the numerical errors
of less than 1\%~\cite{Dic01}.

In fact it was pointed out many years ago~\cite{Alt72} that the GDH sum rule
should hold at leading order in perturbation theory in the standard model of
electroweak interactions. Later on this result was generalized to any
$2\rightarrow2$ process in supersymmetric and other field
theories~\cite{Bro95}. The essential criterion is, of course, that these
``fundamental'' theories start from pointlike particles, and then the GDH sum
rule should hold order by order in the coupling constant. In passing we note
that the GDH sum rule was also investigated in quantum gravity to one-loop
order~\cite{Gol00}. The result was a violation of this sum rule, but that may
be due to our ignorance of quantum gravity in the strong coupling (high energy)
region.

The situation is quite different in effective field theories such as
ChPT~\cite{JiO01}. In this case an unsubtracted dispersion relation like
Eq.~(\ref{e2kappa2}) will not work, because the theory freezes out the
high-energy degrees of freedom in terms of LECs. We may summarize the situation
as follows: In a ``fundamental'' theory such as QED, we can verify
Eq.~(\ref{e2kappa2}) order by order in the coupling constant, and if the
integral on the RHS converges, it must converge to the value of the LHS to that
order. On the other hand, an ``effective'' theory such as ChPT cannot decide
whether the integral exists (in fact it will diverge at any given order).
Instead we must keep subtracting the dispersion relation leading to
Eq.~(\ref{e2kappa2}) until it converges, and since the values at the
subtraction points have to be expressed by LECs like the a.m.m., we lose some
predictive power.

As discussed in Section~2, the discovery of the large a.m.m. of the proton
indicated that this ``elementary'' particle was a microcosm in itself, and in
this sense the findings of Stern and collaborators were revolutionary. In the
following decades the understanding of the nucleon's spin structure underwent
quite a few crises, and ab initio calculations of the nucleon's a.m.m. have
become possible only recently.

In Section~3 we derive the GDH sum rule and related integrals from forward
dispersion relations and low energy theorems for Compton scattering. As the
result of recent experiments, the forward spin polarizability (FSP) of the
proton is now known, and the GDH integral is found to agree with the sum rule
value within the experimental error bars of $\sim10\%$. Data have also been
taken for the neutron and are under evaluation. With some reservations, these
integrals will be generalized to the scattering of virtual photons in
Section~4. The imaginary parts of the respective integrals are related to the
spin-dependent cross sections of electro-excitation and, in the limit of deep
inelastic scattering (DIS), to the spin structure functions. The resulting
generalized integrals and polarizabilities depend on the photon's four-momentum
and therefore contain information on the spatial distribution of the spin
observables. Various GDH-like integrals, the sum rules of Bjorken~\cite{Bjo66},
Burkhardt-Cottingham~\cite{BC70}, and Wandzura-Wilczek~\cite{WW77}, the spin
polarizabilities, and higher twist expansions are discussed and compared with
the experimental data. We conclude in Section~5 with a summary and an outlook
on further developments.

\section*{2.\quad THE ANOMALOUS MAGNETIC MOMENT OF THE NUCLEON}

The a.m.m. $\kappa_p$ of the proton was discovered by Otto Stern and
collaborators~\cite{Ste33} in 1933. For the total magnetic moment $\mu_p$ they
found ``a value between 2 and 3 nuclear magnetons (not 1 as has been previously
expected)''. Their findings came as a big surprise to the community, which had
just learned that Dirac's theory could describe the Bohr magneton of the
electron so well~\cite{Dre00}. The experiment became possible by a continuous
improvement of the molecular beam method, with which
Stern~\&~Gerlach~\cite{Ste21} had found the spin of the electron, the first
indication that elementary particles may carry internal quantum numbers. Though
not immediately recognized as such, the new discovery was even more
revolutionary: A particle with an a.m.m. cannot be a point particle but must
have a finite size, and at the same time the finite size requires an excitation
spectrum of the particle and vice versa.

Shortly after the Yukawa particle had been postulated, Fr\"ohlich et
al.~\cite{Fro38} tried to explain the value of $\kappa$ by a pion cloud. In the
1950's such calculations were even extended to the two-loop level~\cite{Nak50},
but the results were discouraging. When the constituent quark
model~\cite{Mor65} was postulated in the 1960's, it seemed to explain the
a.m.m. right away on the basis of
SU(3)$_{\rm{color}}\times$SU(6)$_{\rm{spin-flavor}}$. In particular it
predicted $\kappa_p/\kappa_n=-1$ for the proton to neutron ratio, quite close
to the experimental value of -0.937. However, this simple picture too had to be
discarded. The analysis of deep inelastic lepton scattering at
SLAC~\cite{Bau83} and CERN~\cite{Ash89} led to the ``spin crisis'' and taught
us that less than half of the nucleon's spin is carried by the quark spins. But
what carries the spin of the nucleon? On the experimental side, semi-inclusive
reactions are expected to answer this question. In particular, deeply virtual
Compton scattering or meson production allows us to define generalized parton
distributions that hold the promise to disentangle the contributions of quark
spin, gluon spin, and orbital angular momentum to the total spin of the
nucleon~\cite{Fil01}.

\begin{figure}
\centerline{\psfig{figure=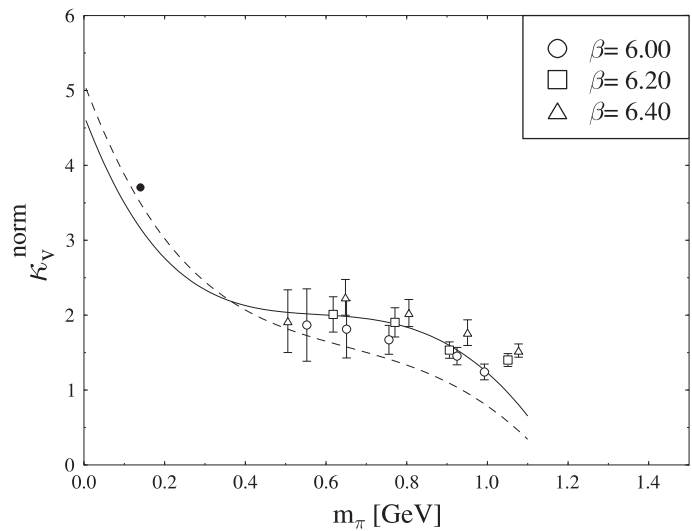,width=9cm}} \caption{Lattice
calculations for the isovector part of the a.m.m., $\kappa_V=\kappa_p-\kappa_n$
(experimental value: 3.71) as a function of the pion mass $m_{\pi}$. Full and
dashed curves: predictions of ChPT with low-energy constants fitted to the
lattice results. See Ref.~\cite{Goe03} for further details \label{fig1}.}
\end{figure}

As far as theory is concerned, a combination of lattice calculations and chiral
field theory is the most promising approach~\cite{Goe03} for ab initio
calculations on the basis of QCD. Unfortunately, the existing lattice
calculations can handle neither the full configuration space (``quenched
calculations'') nor realistic pion masses. As shown in Fig.~\ref{fig1}, the
isovector magnetic moment $\kappa_V$ is evaluated for fictitious pion masses of
500-1100~MeV. These lattice results are then compared to the prediction of ChPT
as a function of the pion mass,
\be \label{kappa_V} \kappa_V = \kappa_V^0-\frac{g_A^2m_{\pi}M}{4\pi f_{\pi}^2} + \ldots \
, \ee
where $g_A=1.267$ is the axial coupling constant and $f_{\pi}=92.4$~MeV is the
pion decay constant. Unfortunately, the absolute value in the chiral limit
$(m_{\pi}\rightarrow 0)$ cannot be predicted but is determined by the LEC
$\kappa_V^0$, which explains the failure of the early attempts to describe the
magnetic moments by pion loops. However, the steep downward slope at
$m_{\pi}=0$ depends only on well-known parameters and is therefore a solid
prediction of ChPT. The ellipses in Eq.~(\ref{kappa_V}) denote higher order
terms, involving in particular two more LECs, which are fitted to the lattice
results. Although the matching of the two schemes over a large range of pion
masses is not without problems, the figure shows that

\begin{itemize}
\item
for pion masses above 400~MeV $\kappa_{V}$ reaches only $\sim50\%$ of its
experimental value, independent of the exact value of $m_{\pi}$, and
\item
the experimental value $\kappa_{V}=\kappa_p-\kappa_n=3.71$ is obtained if the
pion mass approaches its experimental value, $m_{\pi}=140$~MeV, i.e., if the
pion increasingly assumes the properties of a Goldstone boson.
\end{itemize}

Therefore, a simple qualitative picture emerges: About half of the a.m.m.
results from degrees of freedom of the quark core, the other half from the pion
cloud. However, a more quantitative separation of these long- and short-range
phenomena requires a detailed study of their dependence on the ChPT
regularization scale~\cite{BHM04}.

%

\section*{3.\quad COMPTON SCATTERING OF REAL PHOTONS AND THE HELICITY STRUCTURE OF PHOTOABSORPTION}

\subsection*{3.1\quad Forward Dispersion Relations, Low Energy Theorems, and Sum Rules}

In this section we discuss the forward scattering of a real photon by a
nucleon. The incident photon is characterized by the Lorentz vectors of
momentum, $q=(q_0,\,\mathbf{q})$ and polarization,
$\varepsilon_{\lambda}=(0,\,\boldsymbol{\varepsilon}_{\lambda})$, with $q\cdot
q=0$ (real photon) and $\varepsilon_{\lambda}\cdot q=0$ (transverse gauge). If
the photon moves in the direction of the $z$-axis,
${\mathbf{q}}=q_0\,{\hate}_z$, the two polarization vectors may be taken as
\beqn \label{DDeq2.2.1} \bvare_{\pm} = \mp \frac{1}{\sqrt{2}}\,(\hate_x\pm i\hate_y)\ , \eeqn
corresponding to circularly polarized light with helicities $\lambda=+1$
(right-handed) and $\lambda=-1$ (left-handed). The kinematics of the outgoing
photon is then described by the corresponding primed quantities. For the
purpose of dispersion relations we choose the lab frame, and introduce the
notation $q_0^{\mbox{\sc{lab}}}=\nu$ for the photon energy. The absorption of
this photon leads to an excited hadronic system with total c.m. energy
$W=\sqrt{M^2+2M\nu}$.

\begin{figure}
\centerline{\psfig{figure=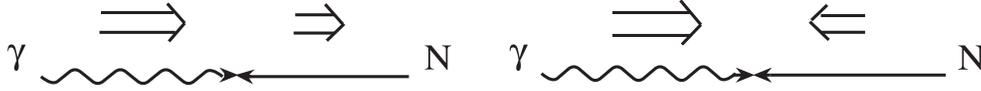,width=13cm}} \caption{Spin and helicity of a
double polarization experiment $\vec{\gamma}+\vec{N}\rightarrow N^{\ast}$. The open
arrows denote the projections of the spin, $S_z$, on the photon momentum, the helicities
$h$ are defined as projections on the respective particle momentum, and the photon is
assumed to be right-handed (helicity $\lambda=+1$):
\newline
left: $N(S_z=1/2,\ h=-1/2)\rightarrow N^{\ast}(S_z=h=3/2)$,
\newline
right: $N(S_z=-1/2,\ h=1/2)\rightarrow N^{\ast}(S_z=h=1/2)$. \label{fig3.1}}
\end{figure}

The forward Compton amplitude takes the general form
\beqn
\label{DDeq2.2.2}
 T(\nu,\,\theta=0) = \bvare'^{\ast}\cdot\bvare\,f(\nu)+
i\,\bsig\cdot(\bvare'^{\ast}\times\bvare)\,g(\nu)\ .
\eeqn
This amplitude is invariant under the crossing transformation,
$\varepsilon'^{\ast}\leftrightarrow\varepsilon$ and $\nu\rightarrow-\nu$, and
therefore $f$ has to be an even and $g$ an odd function of $\nu$. The
amplitudes $f$ and $g$ can be determined by scattering circularly polarized
photons (e.g., helicity $\lambda=1$) off nucleons polarized along or opposite
to the photon momentum $\bq$ as shown schematically in Fig.~\ref{fig3.1}. If
the spins are parallel, the helicity changes by two units and the intermediate
state has helicity $3/2$. Since this requires a total spin $S\ge3/2$, the
transition can only take place on a correlated three-quark system. The case of
opposite spins, on the other hand, is helicity conserving and therefore can
take place on an individual quark. Denoting the Compton scattering amplitudes
for these two experiments by $T_{3/2}$ and $T_{1/2}$, we find
$f(\nu)=(T_{1/2}+T_{3/2})/2$ and $g(\nu)=(T_{1/2}-T_{3/2})/2$. In a similar way
we define the total absorption cross section as the spin average over the two
helicity cross sections,
\beqn \label{DDeq2.2.4} \sigma_T=\frac{1}{2}\,(\sigma_{1/2}+\sigma_{3/2})\ ,
\eeqn
and the transverse-transverse interference term by the helicity difference,
\beqn \label{DDeq2.2.5} \sigma_{TT} = \frac{1}{2}\,(\sigma_{1/2}-\sigma_{3/2})\ . \eeqn
The unitarity of the scattering matrix relates the absorption cross sections to the
imaginary parts of the respective forward scattering amplitudes by the optical theorem,
\begin{eqnarray}
\mbox{Im}\ f(\nu) & = & \frac{\nu}{8\pi} (\sigma_{1/2}(\nu)+\sigma_{3/2}(\nu)) =
\frac{\nu}{4\pi}\,\sigma_T (\nu)\ , \nonumber \\
\mbox{Im}\ g(\nu) & = & \frac{\nu}{8\pi} (\sigma_{1/2}(\nu)-\sigma_{3/2}(\nu)) =
\frac{\nu}{4\pi}\,\sigma_{TT} (\nu)\ . \label{DDeq2.2.6}
\end{eqnarray}

Due to the smallness of the fine structure constant $\alpha_{fs}$ we may
neglect all purely electromagnetic processes, and shall consider only
photoabsorption due to the hadronic channels starting at pion production
threshold, $\nu_0 =m_{\pi}(1+m_{\pi}/2M)\approx150$~MeV. In order to set up the
dispersion integrals, we have to study the behavior of the absorption cross
sections for large energies. The total cross section $\sigma_T$ is essentially
constant above the resonance region, with a slow logarithmic increase at the
highest energies, and therefore we must subtract the dispersion relation for
$f$. If we subtract at $\nu=0$, we also remove the nucleon pole terms at this
point. By use of the crossing relation and the optical theorem, the subtracted
dispersion relation takes the form,
\beqn \label{DDeq2.2.8} {\mbox{Re}}\ f(\nu) = f(0)+\frac{\nu^2}{2\pi^2}\,{\mathcal{P}}\,
\int_{\nu_0}^{\infty}\,\frac{\sigma_T(\nu')} {\nu'^2-\nu^2}\,d\nu'\ . \eeqn
For the odd function $g(\nu)$ we may expect the existence of an unsubtracted dispersion
relation,
\beqn \label{DDeq2.2.9} {\mbox{Re}}\ g(\nu) = \frac{\nu}{4\pi^2}\,{\mathcal{P}}\,
\int_{\nu_0}^{\infty}\,\frac{\sigma_{1/2}(\nu')-\sigma_{3/2}(\nu')}
{\nu'^2-\nu^2}\,\nu'd\nu'\ . \eeqn
If these integrals exist, they can be expanded in a Taylor series at the origin, which
converges up to the lowest threshold, $\nu=\nu_0$~:
\begin{eqnarray}
{\mbox{Re}}\ f(\nu) \,&=&\, f(0)
 + \sum_{n=1}
\left (\frac{1}{2\pi^2}\,\int_{\nu_0}^{\infty}\,
\frac{\sigma_T(\nu')}{\nu'^{2n}}\,d\nu'\right )\,\nu^{2n} \, ,
\label{DDeq2.2.10} \\
{\mbox{Re}}\ g(\nu) \,&=&\,
 \sum_{n=1}
\left (\frac{1}{4\pi^2}\,\int\,
\frac{\sigma_{1/2}(\nu')-\sigma_{3/2}(\nu')}{(\nu')^{2n-1}} \,d\nu' \right )
\,\nu^{2n-1}\ . \label{DDeq2.2.11}
\end{eqnarray}
The expansion coefficients in brackets parameterize the electromagnetic response of the
nucleon.

The results of dispersion theory can be compared to the results of the low
energy theorem (LET) of Low~\cite{Low54}, and
Gell-Mann~\&~Goldberger~\cite{Gel54}. These authors showed that the leading and
next-to-leading terms of the expansions are fixed by the
 Born terms, which can be expressed by the global
properties of the system, the mass $M$, the charge $e\,e_N$, and the a.m.m. $\kappa_N$ of
the respective nucleon ($e_p=1,\ e_n=0,\ \kappa_p=1.79,\ \kappa_n=-1.91)$. Based on
Lorentz invariance, gauge invariance and crossing symmetry, the LET yields the result
\begin{eqnarray}
f(\nu) & = & -\frac{e^2\,e_N^2}{4\pi M} + (\alpha+\beta)
\,\nu^2+ {\mathcal{O}}(\nu^4) \ , \label{DDeq2.2.12} \\
g(\nu) & = & -\frac{e^2\kappa^2_N}{8\pi M^2}\,\nu + \gamma_0\nu^3 + {\mathcal{O}}(\nu^5)
\ . \label{DDeq2.2.13}
\end{eqnarray}
The leading term of the spin-independent amplitude, $f(0)$, is the Thomson term
familiar from nonrelativistic theory, and the term ${\mathcal{O}}(\nu)$
vanishes because of crossing symmetry. The term ${\mathcal{O}}(\nu^2)$
describes Rayleigh scattering and yields information on the internal nucleon
structure through the electric ($\alpha$) and magnetic ($\beta$) dipole
polarizabilities, and the higher order terms ${\mathcal{O}}(\nu^4)$ contain
contributions of dipole retardation and higher multipoles. In the case of the
spin-flip amplitude $g$, the leading term is determined by the a.m.m., and the
term ${\mathcal{O}}(\nu^3)$ contains information on the spin structure through
the forward spin polarizability (FSP) $\gamma_0$. By comparing with
Eqs.~(\ref{DDeq2.2.10}) and (\ref{DDeq2.2.11}), we can construct all higher
coefficients of the low energy expansion, Eqs.~(\ref{DDeq2.2.12}) and
(\ref{DDeq2.2.13}), from moments of the helicity cross sections. In particular
we obtain Baldin's sum rule~\cite{Bal60},
\beqn \label{DDeq2.2.14} \alpha + \beta = \frac{1}{2\pi^2}\,
\int_{\nu_0}^{\infty}\,\frac{\sigma_T(\nu')}{\nu'^2} \,d\nu'\ , \eeqn
the GDH sum rule~\cite{Ger65},
\beqn \label{DDeq2.2.15} \frac{\pi e^2\kappa^2_N}{2M^2}=
\int_{\nu_0}^{\infty}\,\frac{\sigma_{3/2}(\nu') -\sigma_{1/2}(\nu')}{\nu'}\,d\nu' \,
\equiv I \, , \eeqn
and the FSP~\cite{Gel54,GGT54},
\beqn \label{DDeq2.2.16} \gamma_0= \,-\,\frac{1}{4\pi^2}\,\int_{\nu_0}^{\infty}\,
\frac{\sigma_{3/2}(\nu')-\sigma_{1/2}(\nu')} {\nu'^3}\,d\nu'\ . \eeqn
%


\subsection*{3.2\quad Photoabsorption Cross Sections for the Proton}

\begin{figure}
\centerline{\psfig{figure=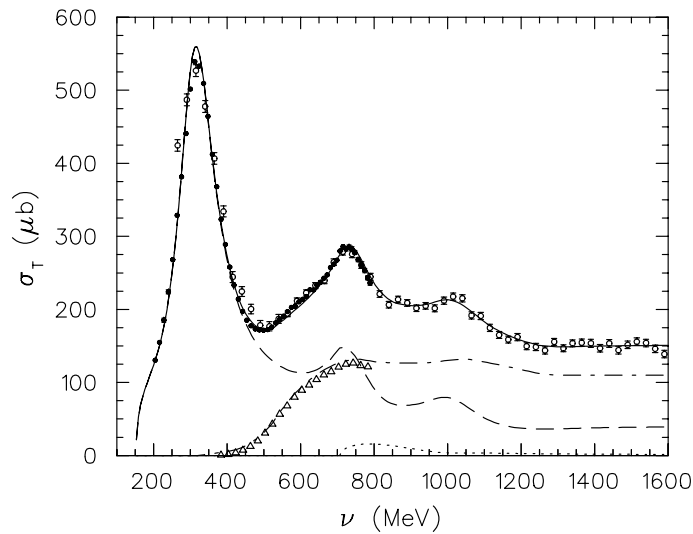,width=9cm,angle=90}} \caption{The total
absorption cross section $\sigma_T$ for the proton. The various lines represent
the MAID results~\cite{Maid03} for the total cross section (solid line),
one-pion channels (dashed line), more-pion channels (dash-dotted line), and
$\eta$ channel (dotted line). The data for the total cross section are from
MAMI~\cite{Mac96} (full circles) and Daresbury~\cite{Dar72} (open circles). The
triangles represent the data for the $2\pi$ channels~\cite{Hol01}
\label{DDfig2.2.2}.}
\end{figure}

The total photoabsorption cross section $\sigma_T$ for the proton is shown in
Fig.~\ref{DDfig2.2.2}. It clearly exhibits three resonance structures on top of
a strong background. These structures correspond, in order, to concentrations
of magnetic dipole strength $(M1)$ in the region of the $\Delta (1232)$
resonance, electric dipole strength $(E1)$ near the resonances $N^{\ast}(1520)$
and $N^{\ast}(1535)$, and electric quadrupole $(E2)$ strength near the
$N^{\ast}(1680)$. For energies above the resonance region $(\nu\gtrsim
1.67$~GeV or $W \gtrsim 2 $~GeV), $\sigma_T$ is slowly decreasing towards a
minimum of $\sim115~\mu$b at $W\approx 10 $~GeV. At the highest energies, $W
\approx 200$~GeV (corresponding to $\nu \simeq 2 \cdot 10^4$~GeV), the
experiments~\cite{DESY} show an increase with energy of the form $\sigma_T \sim
W^{0.2}$, in accordance with Regge parametrizations through a soft pomeron
exchange mechanism~\cite{Cud00}. Although this increase is not expected to
continue forever, we cannot expect that the unweighted integral over $\sigma_T$
converges, and therefore the dispersion relation for the spin-independent
Compton amplitude $f$ has to be subtracted (see Eq.~\ref{DDeq2.2.8}). In
addition to the total cross section, Fig.~\ref{DDfig2.2.2} also shows the
contributions of the most important channels. The one-pion channels dominate up
to $\nu\approx500$~MeV, but in the second resonance region
($\nu\approx700$~MeV) the two-pion channels become quite comparable. The figure
also shows the small contribution of $\eta$ production above
$\nu\approx700$~MeV.

\begin{figure}
\centerline{\psfig{figure=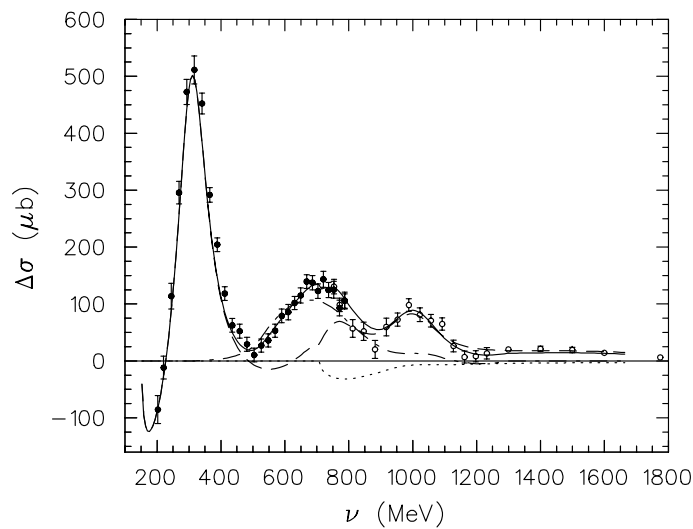,width=9cm,angle=90}} \caption{The helicity
difference $\Delta\sigma = \sigma_{3/2}-\sigma_{1/2}$ for the proton. The
various lines represent the results of MAID (for notation see
Fig.~\ref{DDfig2.2.2}). The experimental data are from MAMI~\cite{Ahr00} (full
circles) and ELSA~\cite{Dut03} (open circles).} \label{DDfig2.2.3}
\end{figure}

Recently, the helicity difference $\sigma_{TT}$ has also been measured. The
first measurement was carried out at MAMI (Mainz) for photon energies in the
range of 200~MeV$ < \nu < 800~$MeV~\cite{Ahr00}, and the GDH Collaboration has
now extended the measurement into the energy range up to 2.9~GeV at ELSA
(Bonn)~\cite{Dut03}. Further experimental activities in this range have been
reported by groups at LEGS (Brookhaven)~\cite{San02}, GRAAL
(Grenoble)~\cite{Bou02}, and SPring-8 (Hyogo)~\cite{Iwa02}. As shown in
Fig.~\ref{DDfig2.2.3}, the helicity difference fluctuates much more strongly
than the total cross section $\sigma_T$. The threshold region is dominated by
s-wave pion production, i.e., intermediate states with spin $1/2$ that can only
contribute to the cross section $\sigma_{1/2}$. In the region of the $\Delta
(1232)$ with spin $J=3/2$, both helicity cross sections contribute, but since
the transition is essentially $M1$, we find
$\sigma_{3/2}/\sigma_{1/2}\approx3$, and the helicity difference
$\Delta\sigma=\sigma_{3/2}-\sigma_{1/2}=-2\sigma_{TT}$ becomes large and
positive. The figure also shows that $\sigma_{3/2}$ dominates the proton
photoabsorption cross section in the second and third resonance regions. In
fact, one of the early successes of the quark model was to predict this feature
by a cancellation of the convection and spin currents in the case of
$\sigma_{1/2}$~\cite{Cop69}.
%
The preliminary data at the higher energies show some indication for the fourth
resonance region (1800~MeV$<W<2000$~MeV) followed by a continuing decrease of
$\Delta\sigma$ and a possible cross-over to negative values at $\nu\gtrsim
2.0$~GeV, as predicted from an extrapolation of DIS data~\cite{Bia99,Sim02}. At
high $\nu$, above the resonance region, one usually invokes Regge phenomenology
to argue that the integral converges~\cite{Bass97}. In particular, for the
isovector channel $\sigma_{1/2} - \sigma_{3/2}$ behaves as $\nu^{\alpha_V - 1}$
at large $\nu$, with $-0.5 \lesssim \alpha_V \lesssim 0$ being the intercept of
the $a_1(1260)$ meson Regge trajectory. For the isoscalar channel, Regge theory
predicts a similar energy behavior with $\alpha_S \simeq - 0.5$, which is the
intercept of the isoscalar $f_1(1285)$ and $f_1(1420)$ Regge trajectories.
However, these ideas should be tested experimentally.

We observe that the large background of non-resonant photoproduction in the
total cross section $\sigma_T$ (Fig.~\ref{DDfig2.2.2}) has almost disappeared
in the helicity difference $\Delta\sigma$ (Fig.~\ref{DDfig2.2.3}), i.e., the
background is ``helicity blind''. This is quite different from the behavior in
the region of DIS, where helicity $3/2$ contributions should vanish relative to
helicity $1/2$ contributions, because only the latter can be produced by
incoherent scattering off a single quark. Obviously the transition from virtual
photons to real photons can not be described by a simple extrapolation. The two
helicity cross sections for real photons remain large and roughly equal up to
the highest energies, at values of
$\sigma_{1/2}\approx\sigma_{3/2}\approx120~\mu$b. Therefore we conclude that
the real photon is essentially absorbed by coherent processes. This absorption
requires interactions among the constituents such as gluon exchange between two
quarks.

\subsection*{3.2\quad The GDH Sum Rule for the Proton}

The GDH sum rule predicts that the integral in Eq.~(\ref{DDeq2.2.15}) takes the
value $I_p=204.8~\mu$b for the proton. As discussed above, the GDH sum rule is
based on three very general principles (Lorentz and gauge invariance and
unitarity) and on one weak assumption: the convergence of an unsubtracted
dispersion relation. Although this assumption can never be proven or disproven
by experiment, it is legitimate to ask whether the a.m.m. on the LHS of
Eq.~(\ref{DDeq2.2.15}) is approximately obtained by integrating the measured
cross sections up to some reasonable energy, say 3 or 50~GeV. This comparison
will tell us whether the a.m.m. measured as a ground state expectation value,
is related to the degrees of freedom visible to that energy, or whether it is
produced by very short-distance and possibly still unknown phenomena, as we
discussed in Section~1.

Let us therefore take a closer look at the different energy regions and their
contributions to the GDH sum rule:
\begin{itemize}
\item
The contributions for $0\le\nu\le\nu_0$, below the pion threshold, are only of
electromagnetic origin and therefore of relative order $\kappa_{\rm{QED}} / \kappa_p \sim
10^{-3}$, where $\kappa_{\rm{QCD}} = \alpha_{fs} / (2 \pi)$ is the Schwinger correction
to the a.m.m. of the proton.
\item
The GDH experiments at MAMI and ELSA used longitudinally polarized protons
provided by a frozen-spin butanol (C$_4$H$_9$OH) target. Although the
(spinless) C and O atoms contained in this target should not affect the
helicity difference, they give rise to large backgrounds for $\sigma_{1/2}$ and
$\sigma_{3/2}$ separately. In order to obtain a good statistical accuracy, it
is therefore a prerequisite to make sure the event happened on a target proton
by detecting the emitted pions or the recoil proton under the right kinematical
conditions. Of course such an experiment becomes increasingly difficult in the
threshold region of $\nu<200$~MeV. However, the absorption cross sections in
that region are completely dominated by the s- and p-wave multipoles, which are
known from experimental analysis and are well described by ChPT and dispersion
theory.
\item
The energy range of the MAMI experiment, 200 MeV$\le\nu\le800$~MeV, includes
the first and part of the second resonance region. The GDH Collaboration has
extended the measurement of the helicity cross sections up to energies of
$\nu=2.9$~GeV at ELSA, thus covering the full resonance region and very
probably the onset of the Regge regime.
\item
The approved SLAC experiment E-159~\cite{E159} will measure the helicity
difference $\Delta\sigma$ for protons and neutrons in the photon energy range
of 5~GeV $< \nu < 40$~GeV. The energy region of $2.5$~GeV$<\nu<5.5$~GeV was
covered in a short run by the CLAS Collaboration at JLab~\cite{Sob02}. The
preliminary data indeed show a negative contribution of $\sim4~\mu$b over that
range, although this is much smaller than previously expected. The SLAC
experiments will test the convergence of the GDH sum rule and provide a
baseline for our understanding of soft Regge physics in the spin-dependent
forward Compton amplitude.
\end{itemize}

Table~\ref{tab1} summarizes the results for the proton by showing predictions
for the GDH integral and the FSP. The threshold contribution is evaluated by a
multipole analysis of pion photoproduction~\cite{Maid03}, with an error
estimated by comparing to other analyses~\cite{Arn02}. The resonance region up
to $\nu=2.9$~GeV is determined by the experimental data taken at
MAMI~\cite{Ahr00} and ELSA~\cite{Sch03}, and the asymptotic region is estimated
from two different Regge analyses~\cite{Bia99,Sim02}. Summing up these
contributions, the GDH sum rule value is obtained within the error bars.
Assuming that the estimated size of the Regge contribution will be confirmed by
the proposed SLAC experiment E159, one may conclude that the GDH sum rule works
for the proton. Because of the different energy weighting with $\nu$, the FSP
integral converges much better and is therefore completely determined by the
energy range of the existing data.

\begin{table}
\def~{\hphantom{0}}
\caption{The contribution of various energy regions to the GDH integral $I$ and
the forward spin polarizability $\gamma_0$ of the proton (see text for
explanation).
\label{tab1}}
\begin{tabular}{@{}lcc@{}}
\toprule
energy range & $I_p$ $[\mu$b]  & $\gamma_0^p$ $[10^{-4}$~fm$^4]$ \\
\colrule
$\nu_0\le\nu\le200$~MeV~\cite{Maid03,Arn02}  &  $-28.5\pm2$  &  $0.95\pm0.05$\\
200 MeV$\le\nu\le800$~MeV~\cite{Ahr00}  &  $226\pm5\pm12$  &  $-1.87\pm0.08\pm0.10$\\
800 MeV$\le\nu\le2.9$~GeV~\cite{Sch03}  &  $27.5\pm2.0\pm1.2$  & $-0.03$  \\
$\nu\ge2.9$~GeV~\cite{Bia99,Sim02}  &  $-14\pm2$  &  $+0.01$ \\
\colrule
total & $211\pm15$  &  $-0.94\pm0.15$ \\
\colrule
sum rule~\cite{Ger65} &  204  &  -- \\
\botrule
\end{tabular}
\end{table}

\subsection*{3.4\quad The Helicity Structure of the Decay Channels}

\begin{figure}
\centerline{\psfig{figure=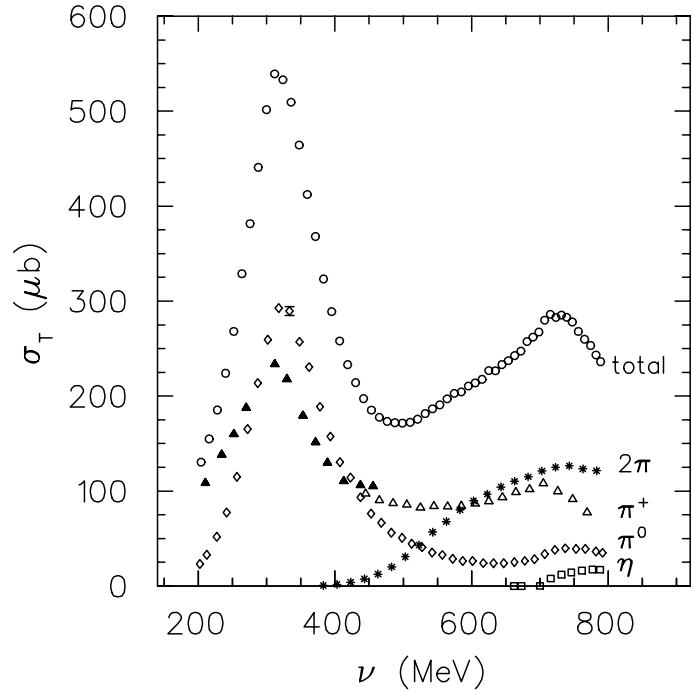,width=9cm}}
\caption{Decay channels for photoabsorption on the proton according
to various MAMI experiments. The data for total photoabsorption (open circles,
Ref.~\cite{Mac96}) are compared to the following partial decay channels: $n\pi^+ $ (solid
and open triangles, Refs.~\cite{Mac96,Ros04}), $p\pi^0$ (diamonds, Ref.~\cite{Har96}),
two-pion production (asterisks, Refs.~\cite{Hol01}), and $p\eta$ (squares,
Ref.~\cite{Kru95}).} \label{channels}
\end{figure}

The cross sections $\Delta\sigma$ and $\sigma_T$ are divided into the
contributions of the decay channels in Figs.~\ref{DDfig2.2.3} and
\ref{channels}. Up to the first resonance region the one-pion channels yield
essentially all of the cross section. This changes in the second and third
resonance regions, where the branching ratio for one-pion production,
$x_{\pi}=\Gamma_{\pi}/\Gamma_{\rm{tot}}$, decreases to values near and below
50~\%. Most of the remainder comes from the channels $\pi\pi$ and $\eta$. All
other channels, including vector mesons ($\rho,\ \omega$) and strangeness
production ($K\Lambda,\ K\Sigma$) yield only a small fraction of the GDH sum
rule~\cite{Zha02}. Table~\ref{tab2} gives a detailed break-down of the sum rule
into the decay channels. The numbers in this table refer to the resonance
region, $\nu\le1.67$~GeV or $W\le2$~GeV, except for the two-pion channels, for
which no estimates exist above $\nu=800$~MeV. Also shown in Table~\ref{tab2}
are the estimated asymptotic contributions for $\nu>1.67$~GeV. The table shows
that the multipole analysis for one-pion production and the more
phenomenological models for the heavier mass channels are in reasonable
agreement with the sum rule in the case of the proton. However, the same
analysis fails in the case of the neutron, which leaves an unexplained ``gap''
of $\sim50~\mu$b between the model predictions and the sum rule.

\begin{table}
\caption{The contribution of various decay channels to the GDH integral $I$ and
the forward spin polarizability $\gamma_0$. The integration extends to
$\nu_{\rm{max}}=1.67$~GeV ($W_{\rm{max}}=2$~GeV) except that the two-pion
contribution is integrated only up to $\nu_{\rm{max}}=800$~MeV. The last row
shows the results of Regge fits for the region $\nu>1.67$~GeV. }
\label{tab2}
\begin{tabular}{@{}l|lcc|lcc@{}}
\toprule
Ref. & proton & $I_p$  & $\gamma_0^p$  & neutron & $I_n$  & $\gamma_0^n$ \\
\colrule
\cite{Maid03}/\cite{Arn02} & $\pi^0 p$ & 157/142 & -1.46/-1.40 & $\pi^0 n$ & 145/147 & -1.44/-1.44 \\
\cite{Maid03}/\cite{Arn02} & $\pi^+n$ & 7.5/44 & 0.82/0.55 & $\pi^-p$ & -21/-13 & 1.53/1.36 \\
\cite{etaMaid} & $\eta p$ & -9.0 & 0.01 & $\eta n$ & -5.9 & 0.01 \\
\cite{Hir02} & $\pi\pi N$
& $28$ & $-0.07$ & $\pi\pi N$ & $19$ & $-0.05$ \\
\cite{Zha02} & $K\Lambda,\ K\Sigma$ & -4.0 & $<0.01$ &
$K\Lambda,\ K\Sigma$ & 2.0 & $<0.01$ \\
\cite{Zha02} &  $\omega p,\ \rho N$ & -3.0 & $<0.01$ &
$\omega n,\ \rho N$ & 2.1 & $<0.01$ \\
\cite{Bia99}/\cite{Sim02} & Regge & -25/-9 & $<0.01$ & Regge &
31/16 & $<0.01$ \\
\botrule
\end{tabular}
\end{table}

In the MAMI experiment on the proton, all decay channels were separately
identified in the range of 200~MeV$<\nu<$800~MeV. Let us have a closer look at
the contributing channels in order. For further details we refer the reader to
a recent review by Krusche~\&~Schadmand~\cite{Kru03}.

\vspace{0.4cm} \noindent{\bf One-pion channels:} One-pion channels open at
$\nu_0\approx150$~MeV and dominate the cross section up to $\nu\approx500$~MeV
except for small contributions due to photonic decay of the excited states and
the onset of two-pion production. Whereas the GDH integral receives substantial
contributions also from heavier particles, the FSP is almost totally saturated
at $\nu = 800$~MeV. The helicity-dependent cross section for one-pion
production may be expanded in the following multipole series~\cite{Dre01}:
\begin{eqnarray}
\label{sigma'TT} \sigma_{TT} (1 \pi) & = & 4\pi \frac{k_{\pi}} {k_{\gamma}} \sum_l
\frac{1}{2}(l+1) \left[(l+2)(|E_{l+}|^2 + |M_{l+1,-}|^2)\right.
\\ &&
\left.- l(|E_{l+1,-}|^2 + |M_{l+}|^2) + 2l(l+2)\ \mbox{Re}\ (E_{l+}^{\ast}M_{l+}-
     E^{\ast}_{l+1,-}M_{l+1,-})\right]  \nonumber \\
& = & 4\pi \frac{k_{\pi}} {k_{\gamma}}\, (|E_{0+}|^2 - |M_{1+}|^2 + 6\ \mbox{Re}\
(E_{1+}^{\ast}M_{1+}) + 3|E_{1+}|^2 \nonumber
\\ &&
+ |M_{1^-}|^2 - |E_{2-}|^2 - 6\ \mbox{Re}\ (E_{2-}^{\ast}M_{2-}) + 3|M_{2^-}|^2 \pm \ ...
)\ , \nonumber
\end{eqnarray}
where $k_{\pi}$ and $k_{\gamma}$ are the c.m. momenta of pion and photon,
respectively. The electric (E) and magnetic (M) multipoles~\cite{CGLN} in this
expansion are defined by the quantum numbers of the hadronic final state, with
the first index ($\ell$) referring  to the pion-nucleon orbital momentum and
the second index ($\pm$) describing the coupling of orbital angular momentum
and nucleon spin to the total angular momentum of the hadronic state
($J=\ell\pm\textstyle{\frac{1}{2}}$).

As we have seen, the threshold region was not covered by the experiment. The
dominant multipole in this region is $E_{0+}$, which corresponds to an electric
dipole (E1) transition leading to the production of (mostly charged) pions in
an s wave. This multipole is well described by the analysis of (unpolarized)
pion photoproduction in terms of angular distributions, ChPT~\cite{Ber91},
dispersion theory~\cite{Han98}, and phenomenological
analysis~\cite{Maid03,Arn02}. Also the onset of the p-wave multipoles seems to
be well under control~\cite{Ber95}.

With increasing photon energy, the first resonance becomes more and more
dominant, mainly because of the magnetic dipole (M1) excitation of the $\Delta$
(1232), which is described by the multipole $M_{1+}$. The associated electric
multipole $E_{1^+}$ corresponding to E2 excitation is strongly suppressed by
the small ratio $R=E_{1^+}/M_{1^+}=(-2.5\pm0.2\pm0.2)$\% as determined earlier
by angular distributions and photon asymmetries for neutral pion
photoproduction~\cite{Bec97}. Because the product of these multipoles appears
with a factor~6 in Eq.~(\ref{sigma'TT}), the ``GDH experiment'' allows for an
independent measurement~\cite{Ahr04} with the result
$R=(-2.74\pm0.03\pm0.5)\%$. Altogether the MAMI data are in good agreement with
the multipole analyses in the first resonance region~\cite{Maid03,Arn02}, and
also with the preliminary LEGS data~\cite{San02}. However, even relatively
small effects count in this region, because it provides the lion's share to the
sum rule, $I\,(\nu=250-400$~MeV$)=(176\pm8)\,\mu$b!

The finite value of $E_{1^+}$ in the $N\rightarrow\Delta$ transition is
evidence for tensor correlations, which may be attributed to the hyperfine
interaction between the quarks due to gluon exchange. However, the bulk
contribution to $E_{1^+}$ is found to stem from the pion cloud around the quark
bag. From the negative value of $R$ one can derive, in the framework of various
quark models, an oblate deformation of the $\Delta$ resonance.

In a similar way the $N\rightarrow N^{\ast}\,(1520)$ transition was studied by the GDH
Collaboration~\cite{Ahr02D}. The multipoles $E_{2^-}$ and $M_{2^-}$ (E1 and M2
transitions, respectively) are related to the helicity amplitudes,
\be A_{1/2}\,(D_{13})\sim E_{2^-} - 3M_{2^-} \ ,\quad A_{3/2}\,(D_{13})\sim \sqrt{3}
(E_{2^-}+M_{2^-})\ . \ee

The new results~\cite{Ahr02D} are $A_{1/2}=-38\pm3$ (PDG, Ref.~\cite{PDG}:
$-24\pm9$) and $A_{3/2}=147\pm10$ (PDG, Ref.~\cite{PDG}: $166\pm5$), all in
units of $10^{-3}$~GeV$^{-1/2}$. Expressed in terms of the multipoles, the
ratio $M_{2^-}/E_{2^-}$ increases from 0.45 to 0.56, and the respective
one-pion contribution to the sum rule decreases by $\sim25$~\%. From these
examples it should become obvious that double-polarization experiments serve
many purposes besides measuring the GDH integral. In particular, they provide a
very sensitive tool to study resonance properties. For completeness we list the
multipoles (in brackets: the electromagnetic transitions) of some other
resonances. The Roper resonance $N^{\ast}\,(1440)$ with multipole $M_{1^-}(M1)$
contributes with the same sign as s-wave pion production, the
$N^{\ast}\,(1535)$ appears as a resonance in the multipole $E_{0^+}(E1)$ just
above $\eta$ threshold and dominates that decay process, and the most important
resonance in the third resonance region is the $N^{\ast}\,(1680)$ with
multipoles $E_{3^-}(E2)$ and $M_{3^-}(M3)$.

\begin{figure}
\centerline{\psfig{figure=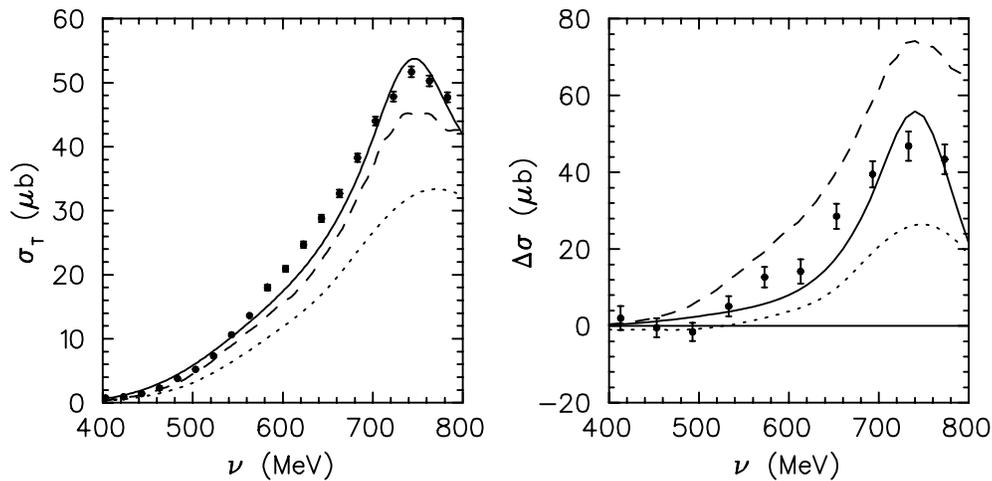,width=13cm,angle=90}} \caption{The total
cross section $\sigma_{T}$ and the helicity difference
$\Delta\sigma=\sigma_{3/2}-\sigma_{1/2}$ for the reaction $\vec{\gamma}
\vec{p}\rightarrow n\pi^+\pi^0$. The theoretical predictions are shown by the
solid lines~\cite{Hir02}, dashed lines~\cite{Nac02}, and dotted
lines~\cite{Hol01a}. The data are from MAMI~\cite{Ahr03pipi}. \label{pipi} }
\end{figure}

\vspace{0.4cm} \noindent{\bf Two-pion channels:} Although the reaction
threshold lies in the $\Delta$ region at $\nu_{2\pi}=309$~MeV, this channel is
practically absent below 400~MeV, and on the scale of Fig.~4 it becomes visible
only for $\nu\gtrsim500$~MeV. The interesting and previously unexpected feature
is the peaking of the respective cross section at $\nu\approx700$~MeV or
$W\approx1480$~MeV, definitely below the positions of the $D_{13}\ (1520)$ and
$S_{11}\ (1535)$ resonances. This is experimental proof that two-pion
production can not be simply explained by a resonance driven mechanism
(s-channel contribution), but that it requires large non-resonant effects such
as Born terms and vector meson exchange in the t-channel. The channels
$n\pi^+\pi^0$, $p\pi^+\pi^-$, and $p\pi^0\pi^0$ were separately analyzed in the
MAMI experiment~\cite{Ahr03pipi}. As an example we show the cross section for
$\vec{\gamma} \vec{p}\rightarrow n\pi^+\pi^0$ in Figure~\ref{pipi}. It leads to
an important contribution of ($11.3\pm0.7\pm0.7)~\mu$b to the GDH integral in
the range $\nu<800$~MeV, whereas its contribution changes the FSP by only
$\sim3$~\%. Figure~\ref{pipi} demonstrates that the present models can only
describe the data in a semi-quantitative way.

\vspace{0.4cm} \noindent{\bf Eta channel:} The $\eta$ channel provides another
important contribution to the sum rule. If it were not for the mixing of the
two particles, the $\eta\ (547)$ would be a member of the pseudoscalar meson
octet and the $\eta'\ (958)$ a member of the corresponding singlet. The latter
acquires its large mass by the $U_A(1)$ anomaly, i.e., by its coupling to the
gluon field. However, because the two particles have the same quantum numbers,
the physical particles are superpositions of the pure octet and singlet states,
and therefore the $\eta$ has a mass much larger than a typical Goldstone boson
like the pion, whose mass would vanish in the limit of zero quark masses. The
threshold for $\eta$ photoproduction is $\nu_{\eta}=706$~MeV. Etas are
essentially produced by the decay of the nearby resonance $S_{11}\ (1535)$,
which has an exceptionally large branching ratio of
$\Gamma_{\eta}/\Gamma_{\rm{tot}}\lesssim50$~\%. This ratio is at most a few
percent for all other resonances. Because the particle is pseudoscalar and the
dominating resonance has $\ell=0$, this channel should essentially contribute
to the helicity cross section $\sigma_{1/2}$ and the multipole $E_{0^+}$. This
prediction was recently confirmed by a direct measurement at
MAMI~\cite{Ahr03eta} in the $S_{11}\ (1535)$ resonance region. The result for
$\sigma_{3/2}$ was compatible with zero, and $\sigma_{1/2}$ came out in perfect
agreement with MAID~\cite{etaMaid}.

\subsection*{3.5\quad The GDH Sum Rule for the Neutron}

Whereas experimental data support the existence of the GDH sum rule for the
proton, the present situation is much less clear in the case of the neutron,
for which the sum rule predicts the value $I_n = 233.2\, \mu b$. From our
knowledge of the CGLN multipoles and models of heavier mass intermediate
states, we obtain ($129\pm5)~\mu$b for the one-pion channel and $-6~\mu$b for
$\eta$ production in the resonance region $W<2$~GeV (see also
Table~\ref{tab2}). An educated guess for two-pion production would be
($40\pm20)~\mu$b, and the asymptotic contribution above 2~GeV has been
estimated to yield another ($25\pm10)~\mu$b. This estimate for $I_n$ falls
short of the sum rule value by $\sim20$~\%. However, given the model
assumptions and the uncertainties in the present data, one certainly cannot
conclude that the neutron sum rule is violated. Possible sources of the
discrepancy may be the neglect of final state interactions for pion production
off the ``neutron targets'' $^2$H or $^3$He, the helicity structure of two-pion
production, or the asymptotic contribution. Some of these open questions should
be answered by the analysis of the data recently taken at ELSA and
MAMI~\cite{GDHneutron} by the GDH collaboration.

\begin{table}
\def~{\hphantom{0}}
\caption{The magnetic moment $\mu$ (in units of $\mu_N$), the a.m.m. $\kappa$, and the
GDH sum ${\rm{rule}}\ I$ (in units of $\mu$b) for electrons, protons, neutrons,
deuterons, and $^3$He nuclei as well as fully polarized hydrogen atoms and butanol
molecules. } \label{kappa}
\begin{tabular}{@{}lccccccc@{}}
\toprule
& $ e $ & $p$  & $n$  & $d$   &  $^3$He & $^1$H & C$_4$H$_9$OH \\
\colrule
$\mu$ & -1838 & 2.793 & -1.913 & 0.857 & -2.128 & -1836 & -1.8$\cdot10^4$  \\
$\kappa$ & $-1.2\cdot10^{-3}$ & 1.793 & -1.916 & -0.143 & -8.371 & -918 & -6.8$\cdot10^4$ \\
$I$  & 289 & 204.8 & 233.2 & 0.65 & 497.8 & 1.1$\cdot10^8$ & 1.1$\cdot10^9$ \\
\botrule
\end{tabular}
\end{table}

Table~\ref{kappa} shows the a.m.m. $\kappa$ for the nucleons and standard
``neutron targets''. The most striking observation is the tiny value of
$\kappa$ for the deuteron, a loosely bound system of a proton and a neutron,
which has isospin $I=0$ and spin $S=1$ and is essentially in a relative s
state. As a consequence the deuteron's magnetic moment $\mu_d$ is roughly the
sum of $\mu_p$ and $\mu_n$, the large isovector component of the nucleon
magnetic moments cancels, and the result is $\mu_d\approx\mu_N$. As can be seen
from Eq.~(\ref{e2kappa2}), the ``normal'' magnetic moment of the deuteron takes
exactly the same value and therefore $\kappa_d\ll1$ and $I_d\ll I_p+I_n$. The
large contributions of the subnuclear degrees of freedom above pion threshold
have to be cancelled by contributions of the nuclear degrees of freedom, and
this must happen to three decimal places!

Let us now apply the same reasoning to $^3$He, a system of two protons with
spins paired off and an ``active'' neutron, essentially again in s states of
relative motion. The result is that $\mu_{^3\rm{He}}\approx\mu_n<0$, whereas,
according to Eq.~(\ref{e2kappa2}), the ``normal'' moment of $^3$He  takes a
positive value. As a consequence $\kappa_{^3\rm{He}}$ has a large negative
value, and $I_{^3\rm{He}}$ is much larger than in the deuteron case.

Arenh\"ovel and collaborators have investigated the interplay between nuclear
and subnuclear degrees of freedom for the deuteron~\cite{Are97}. They begin
with a nonrelativistic potential model approach but include meson exchange and
isobar currents as well as relativistic corrections. The most important channel
is deuteron disintegration, $\gamma+d\rightarrow p+n$, which yields a maximum
value of $\sigma_P-\sigma_A\approx-1800~\mu$b at $\nu\approx2.3$~MeV owing to
the M1 transition $^3S_1\rightarrow{}^1S_0$. Note that this nuclear transition
changes the magnetic moments of the constituents from parallel to antiparallel.
To the contrary, the $N\rightarrow\Delta$ transition leads to a parallel
orientation of all quark spins and peaks at $\nu\approx330$~MeV with a maximum
value of $\sigma_P-\sigma_A\approx+700~\mu$b. As a function of
$\nu_{\rm{max}}$, the upper limit of integration, the disintegration channel
yields the following contributions to the GDH sum rule:
$I_{\rm{pn}}\,(10~\rm{MeV})\approx-600~\mu$b,
$I_{\rm{pn}}\,(150~\rm{MeV})\approx-510~\mu$b, and
$I_{\rm{pn}}\,(550~\rm{MeV})\approx-413~\mu$b. This channel seems to be
saturated at 550~MeV, which also marks the limit of a nonrelativistic
calculation. Coherent pion production gives another important contribution,
$I_{\pi^0d}\,(550~\rm{MeV})=63~\mu$b. In total the calculation yields
$-350~\mu$b from these two coherent processes. The contributions of incoherent
production of pions and heavier mass systems, as estimated from the free
nucleons, yield $I_p+I_n=438~\mu$b. As a result the cancellation of nuclear and
subnuclear contributions is indeed large, but not large enough to reach the
extremely small value of $I_d$.

Eagerly waiting for the results of the ELSA and MAMI experiments on the deuteron target,
we can only speculate about the open questions:

\begin{itemize}
\item
A first glimpse of the reaction
$\vec{\gamma}+\vec{d}\rightarrow\pi^-+p+p_{\rm{spectator}}$ gives the
impression that final state scattering is small, and that the impulse
approximation may be an adequate treatment~\cite{Ped04}. Although this leads to
some reduction in the $\Delta$ region, the overall change of the sum rule
contribution by binding effects is probably small. However, a 20~\% reduction
of the incoherent contribution would be necessary to reach the small value of
$I_d$.
\item
The positive contributions of the two coherent processes around $\nu=300$~MeV
are certainly driven by photo-excitation of the $\Delta$ resonance with the
decay pion being reabsorbed by the spectator nucleon. On the other hand the
nearby possibility of pion photoproduction will also be felt in the
photodisintegration process below threshold. Nuclear and subnuclear degrees of
freedom are thus deeply interwoven.
\item
Relativistic corrections~\cite{Are97} decrease $I_{pn}$ from $-619~\mu$b to $-413~\mu$b,
which may indicate the need for a truly Lorentz invariant description. In spite of such
possible shortcomings, the non-relativistic calculation fulfills an interesting relation
in the limit $\kappa_{p,n}\rightarrow0$: The contribution of deuteron breakup, $I_{pn}$,
tends to zero as is required for consistency, because also the contributions of pion
photoproduction vanish in that limit.
\item
On the basis of the existing calculations, the simple recipe to derive $I_n$
from the inclusive helicity cross sections above pion threshold is bound to
fail. We also run into serious conceptual problems if we somehow subtract the
coherent processes above that threshold or divide them between the nucleons and
the nucleus. The only real hope for a quantitative solution of the neutron
puzzle is that the final data analysis might essentially confirm the validity
of the impulse approximation for incoherent pion production. By correcting for
Fermi motion effects, it may then be possible to derive the helicity cross
sections for the free neutron from the experimental values, with the additional
check that this procedure should also work for the proton.
\end{itemize}

Concerning the $^3$He target, the ongoing electroproduction experiments at JLab
(see Section~4) are presently analyzed within the impulse approximation and
including a (small) depolarization factor due to the probability of relative d
states in $^3$He. Because the proton spins are paired off and the impulse
approximation leads only to small modifications for the polarized neutron, the
inclusive helicity difference above pion threshold is then assumed to yield
$I_n$ modulo the corrections for Fermi motion. As a result the nuclear effects
below pion threshold should yield a contribution of $\sim260~\mu$b in order to
fulfill the sum rule for $^3$He. However, this procedure can only yield a first
estimate, because non-pionic breakup and coherent pion production (${}^3$He\
$\pi^0$ and ${}^3$H\ $\pi^{+}$ final states!) will certainly complicate the
analysis.

As we stated above, the interplay of nuclear and subnuclear degrees of freedom
is based on the general principles that enter the low-energy theorems and
dispersion relations. A complete answer to the questions arising in this
context calls for experiments over the full energy range. It is therefore very
promising that programs are also developing for energies between nuclear
breakup and pion threshold at TUNL/HI$\gamma$S (Duke)~\cite{TUNL}, both for the
nuclear physics aspects by themselves and as a test of many-body calculations
that inevitably will be required to determine whether the GDH sum rule is
fulfilled for the neutron.

We conclude this subsection with a possibly academic but still interesting
question. Although we have talked about nucleons and nuclei as targets, we
could just as well think about projects to measure the GDH sum rule for atoms
or molecules. Table~\ref{kappa} shows the a.m.m. $\kappa$ and the GDH sum rule
$I$ for several such systems. The comparison of the different hierarchies is
quite amusing. If one tried to reconstruct, for example, the a.m.m. of the
hydrogen atom by a GDH integral over its atomic spectrum in the eV or keV
region, one would find physics ``beyond'': the physics of $e^+e^-$ pair
production at the MeV scale (QED) and hadronic physics above pion threshold
(QCD).

\subsection*{3.6\quad Predictions for the Spin Polarizabilities}

Because the GDH sum rule involves the magnetic moment, ChPT can not predict the
sum rule but tacitly assumes its validity by inserting the appropriate LECs.
However, there are genuine predictions for the polarizabilities and in
particular also for the FSP. The first calculation of Compton scattering in
ChPT was performed by Bernard et al. in 1991~\cite{BKM91}. Keeping only the
leading term at order $p^3$ they obtained the result
\be \label{DDeq3.9.4} \alpha = 10\beta = \frac {5\alpha_{fs}g_A^2}{96\pi
f_{\pi}^2m_{\pi}} = 12.2\cdot 10^{-4}~{\rm{fm}}^2\ , \ee
in remarkable agreement with experiment (see Ref.~\cite{Olm01} and references
cited therein). The calculation was later repeated in heavy baryon ChPT, which
allows for a consistent chiral power counting, and extended to
${\mathcal{O}}(p^4)$. At that order there appear several LECs, which were
determined by resonance saturation~\cite{BKSM93}, i.e., by putting in
phenomenological information about the $\Delta$ (1232) resonance. Because this
resonance lies close by, it may not be justifiable to ``freeze'' its degrees of
freedom. It was therefore proposed~\cite{HHK97} to include the excitation
energy of the $\Delta$ (1232) by means of an additional expansion parameter
(``$\varepsilon$ expansion''). Unfortunately, at ${\mathcal{O}}(\varepsilon^3)$
the ``dynamical'' $\Delta$ increases the polarizabilities to values far above
the data. Since large loop corrections are expected at
${\mathcal{O}}(\varepsilon^4)$, a calculation to this order might remedy the
situation. Otherwise one would have to shift the problem to large contributions
of counter terms, thus losing the predictive power.

\begin{table}
\caption{Theoretical predictions for the forward spin polarizability of the
proton and neutron in units of $10^{-4}$ fm$^4$. The table contains the results
of heavy baryon ChPT at ${\mathcal{O}}(p^3)$~\cite{BKKM92} and
${\mathcal{O}}(p^4)$~\cite{Kum00}, the small scale expansion~\cite{HHKK98}, the
full one-loop result (not expanded!) of Lorentz-invariant baryon
ChPT~\cite{Ber03}, fixed-$t$ dispersion relation~\cite{DPV03,Bab98a},
hyperbolic dispersion relations~\cite{DPV03}, a dressed K-matrix
model~\cite{KS01} and the hypercentral quark model~\cite{Gor04}. }
\label{tab4}
\begin{tabular}{lccccccccc}
\toprule Ref. & \cite{BKKM92} & \cite{Kum00} & \cite{HHKK98}
 & \cite{Ber03}& \cite{DPV03} & \cite{Bab98a}& \cite{DPV03}& \cite{KS01}& \cite{Gor04} \\
\colrule
$ \gamma_0^p$  & 4.6  & -3.9   & 2.0   & 4.6  & -0.7   &-1.5  & -1.1 & 2.4 & -1.0 \\
$ \gamma_0^n$  &4.6   &-0.7    & 2.0    & 1.8   & -0.07   & 0.4  & -0.5 & 2.0 & -1.1 \\
\botrule
\end{tabular}
\end{table}

The spin polarizabilities were first predicted in heavy baryon ChPT~\cite{BKKM92} at
${\mathcal{O}}(p^3)$. Two later calculations~\cite{Kum00} at ${\mathcal{O}}(p^4)$ yielded
the result
\begin{eqnarray}
\label{DDeq3.9.8} \gamma_0^p & = & \frac{\alpha_{fs}g_A^2}{24\pi^2f_{\pi}^2m_{\pi}^2}
\left(1-\frac{\pi m_{\pi}}{8M} (21+4\kappa_p-2\kappa_n)\right) \nonumber \\
& = & (4.5-8.3)\cdot 10^{-4}\,fm^{-4} = -3.8\cdot 10^{-4}\,fm^{-4}\ .
\end{eqnarray}

This calculation was recently repeated within a newly developed
Lorentz-invariant formulation of baryon ChPT~\cite{Ber03}. The full one-loop
result, expanded as a power series in $m_{\pi}$, takes the form
$\gamma_0^p=4.5-8.3+6.0+3.2+{\mathcal{O}}(m_{\pi}^2)=4.6$ (in units of
$10^{-4}$~fm$^{-4}$). The leading and subleading terms agree, of course, with
Eq.~(\ref{DDeq3.9.8}). However, the following terms indicate that the expansion
does not seem to converge to the experimental result of Table~\ref{tab2}.
Equation~(\ref{DDeq3.9.8}) shows that the bad convergence is due to
unexpectedly large prefactors of the expansion parameter $m_{\pi}/M\approx1/7$.
As has been pointed out, this problem is essentially due to the Born
terms~\cite{Ber03}. However, we will have to wait for a full two-loop
calculation (and possibly for the inclusion of the ``dynamical'' $\Delta$),
before we can draw definite conclusions about the convergence of the FSP as a
function of $m_{\pi}$.

Table~\ref{tab4} lists several predictions for the FSP. For more details about
the polarizabilities of the nucleon, we refer the reader to Ref.~\cite{DPV03}
and several recent investigations including the $\Delta(1232)$ as an explicit
degree of freedom~\cite{Ber03,Hil03}. Finally we would like to mention that the
first predictions of lattice QCD have recently appeared~\cite{latticepol},
though as yet only for the scalar polarizabilities $\alpha$ and $\beta$ and
pion masses above 550~MeV.

\section*{4.\quad DOUBLY-VIRTUAL COMPTON SCATTERING AND GENERALIZED GDH INTEGRALS}

\subsection*{4.1\quad VVCS and Inclusive Lepton Scattering}

In this section we consider the forward scattering of a virtual photon with
space-like four-momentum $q$, i.e., $q^2=q^2_0-{\bq}^2 =-Q^2<0$. This
doubly-virtual Compton scattering (VVCS) offers a useful framework to study the
generalized GDH integrals and polarizabilities~\cite{Ji93}. However, it is not
merely a theoretical construct. As pointed out many years ago~\cite{Ruj71}, the
imaginary part of this amplitude can be studied by measuring the (transverse)
vector analyzing power or single spin asymmetry (SSA) in elastic lepton
scattering. The SSA is parity conserving but time-reversal odd, and therefore
it vanishes in first-order perturbation theory. As a result the measured
asymmetry is determined by the product of the (real) Born amplitude for
one-photon exchange with the imaginary part of two-photon exchange, which is of
course given by the VVCS tensor. A pioneering experiment to measure the SSA has
been performed at MIT/Bates~\cite{Wel01}, and more data are expected from
ongoing or proposed experiments at the JLab~\cite{Kum99}, SLAC~\cite{Kum04},
and MAMI~\cite{Maa04}. The SSA has also been discussed in the context of deeply
virtual Compton scattering for the transition from a space-like to a time-like
photon, which offers the possibility to study generalized parton
distributions~\cite{Gui03}.

The absorption of a virtual photon on a nucleon $N$, is related to inclusive
electroproduction, $e+N\rightarrow e'+$ anything, where $e$ and $e'$ are the
electrons in the initial and final states, respectively. The kinematics of the
electron is traditionally described in the lab frame (rest frame of $N$), with
$E$ and $E^{'}$ denoting the initial and final energies of the electron,
respectively, and $\theta$ the scattering angle. This defines the kinematics of
the virtual photon in terms of four-momentum transfer $Q$ and energy transfer
$\nu$,
\beqn Q^{2} = 4 EE^{'} \sin^{2} \frac{\theta}{2} , \quad \nu = E - E^{'}\ .
\label{DDeq2.3.1} \eeqn
In the c.m. frame of the hadronic intermediate state, the four-momentum of the
virtual photon is given by
\beqn \omega_{\gamma} = \frac{M \nu - Q^2}{W}, \quad {\bk}_{\gamma} = \frac{M}{W} \,
{\bq}\ , \label{DDeq2.3.2} \eeqn
where $|{\bq}|= \sqrt{Q^2 + \nu^2}$ is the lab photon momentum and $W= \sqrt{2M
\nu + M^{2}-Q^2}$ the total energy in the hadronic c.m. frame. We also
introduce the Bjorken variable $x=Q^2/2M\nu$. The virtual photon spectrum is
normalized according to Hand's definition~\cite{Han63} by the ``equivalent
photon energy'' $K = K_H = \nu (1-x) = (W^{2}-M^{2})/2M$. This replaces the
photon energy, which according to Eq.~(\ref{DDeq2.3.2}) vanishes in the c.m.
frame at $\nu = Q^2/M$ and therefore is inconvenient in the context of the
multipole expansion. However, an alternative would be Gilman's
definition~\cite{Gil68}, $K_G = |{\bq}|$.

The inclusive inelastic cross section may be written in terms of a flux factor and four
partial cross sections~\cite{Dre01},
\beqn \frac{d\sigma}{d\Omega\ dE'} = \Gamma_V \sigma (\nu,Q^2)\ , \label{DDeq2.3.6} \eeqn
\beqn \sigma = \sigma_T+\epsilon\sigma_L-hP_x\sqrt{2\epsilon(1-\epsilon)}\
         \sigma_{LT}-hP_z\sqrt{1-\epsilon^2}\ \sigma_{TT}\ ,
\label{DDeq2.3.7} \eeqn
with the photon polarization $\epsilon$ and the virtual photon flux factor $\Gamma_V$,
\beqn \epsilon = \frac{1}{1+2 (1+ \nu^2/Q^2) \tan^2 \theta/2}\ ,\quad \Gamma_V =
\frac{\alpha_{fs}}{2\pi^2}\ \frac{E'}{E}\ \frac{K}{Q^2}\ \frac{1}{1-\epsilon}\ .
\label{DDeq2.3.8} \eeqn
In addition to the transverse ($\sigma_T$) and longitudinal ($\sigma_L$) cross
sections, the polarization of the virtual photon gives rise to
longitudinal-transverse ($\sigma_{LT}$) and transverse-transverse
($\sigma_{TT}$) interference terms. The two spin-flip (interference) cross
sections can only be measured by a double-polarization experiment, in which $h
= \pm 1$ refers to the two helicity states of the (relativistic) electron, and
$P_z$ and $P_x$ denote the components of the target polarization in the
direction of the virtual photon momentum ${\bq}$ and perpendicular to that
direction in the scattering plane of the electron. Contrary to the sign
convention of Reference~\cite{Dre01}, we have followed the standard notation of
DIS: $\sigma_{LT/TT}$ (Equation~\ref{DDeq2.3.7}) $=-\sigma'_{LT/TT}$
(Reference~\cite{Dre01}).

The spin-flip cross sections are related to the quark structure functions $g_1$ and $g_2$
as follows~\cite{DPV03}:
\begin{equation}
\label{DDeq2.3.11} \sigma_{TT} = \displaystyle{\frac{4\pi^2\alpha_{fs}}{M K}}
\left(g_1-\gamma^2\, g_2 \right)\ ,\quad \sigma_{LT} =
\displaystyle{\frac{4\pi^2\alpha_{fs}}{M K}}\,\gamma\,\left(g_1+g_2\right)\ ,
\end{equation}
with $\gamma = Q/\nu$. The one-pion contribution to these cross sections may be expressed
in terms of the electric (E), magnetic (M), and Coulomb or ``scalar'' (S) multipoles,
\begin{eqnarray}
\label{sigmaTT} \sigma_{TT}^{(1 \pi)} & = & 4\pi \frac{k_{\pi}} {K_{W}} \sum_l
\frac{1}{2}(l+1) \left[(l+2)(|E_{l+}|^2 + |M_{l+1,-}|^2)\right.
\\ && \left. - l(|E_{l+1,-}|^2 + |M_{l+}|^2) + 2l(l+2)
{\rm{Re}} (E_{l+}^{\ast}M_{l+}-
     E^{\ast}_{l+1,-}M_{l+1,-})\right]  \nonumber \ ,
\end{eqnarray}
\begin{eqnarray}
\label{sigmaLT} \sigma_{LT}^{(1 \pi)} & = & 4\pi \frac{k_{\pi}} {K_W}
\frac{Q}{k_{\gamma}}
\sum_l\frac{1}{2}(l+1)^2\\
&& \cdot {\rm{Re}} \left[S_{l+}^{\ast}((l+2)E_{l+} + lM_{l+}) + S_{l+1,-}^{\ast}
   (lE_{l+1,-} - (l+2)M_{l+1,-})\right] \nonumber \ ,
\end{eqnarray}
with $K_W = \frac{M}{W}K$, and $k_{\pi}$ and $k_{\gamma}$ the c.m. momenta of
pion and photon, respectively.

The VVCS amplitude for forward scattering of virtual photons generalizes
Eqs.~(\ref{DDeq2.2.1}) and (\ref{DDeq2.2.2}) by introducing an additional longitudinal
polarization vector $\hat{\bq}$,
\begin{eqnarray}
\label{DDeq2.3.19} T(\nu,\,Q^2,\,\theta=0)& \;=\; & {\bvare}\,'^{\ast}\cdot{\bvare} \,
f_T(\nu,\,Q^2) \;+\;
f_L(\nu,\,Q^2) \nonumber\\
&+& \;i{\bsig}\cdot({\bvare}\,'^{\ast}\times{\bvare}) \,g_{TT}(\nu,\,Q^2)
\nonumber\\
&+& i\,({\bvare}\,'^{\ast}-{\bvare})\cdot({\bsig} \times \hat{\bq} \,)
\,g_{LT}(\nu,\,Q^2) \ .
\end{eqnarray}

In order to construct the VVCS amplitudes in one-to-one correspondence with the
nucleon structure functions, it is useful to cast Eq.~(\ref{DDeq2.3.19}) in a
covariant form,
\begin{eqnarray}
\label{DDeq2.3.48} T(\nu,\,Q^2,\,\theta=0) \;=\;
\varepsilon_{\mu}'^{\ast}\varepsilon_{\nu} \hspace{-0.5cm}&&\left \{ \left(
-g^{\mu\nu}+\frac{q^{\mu}q^{\nu}}{q^2}\right)
T_1(\nu,\,Q^2) \right .\nonumber \\
&& \left . +\frac{1}{p\cdot q} \left(p^{\mu}-\frac{p\cdot q}{q^2}\,q^{\mu}\right)
\left(p^{\nu}-\frac{p\cdot
q}{q^2}\, q^{\nu} \right) T_2(\nu,\,Q^2) \right .\nonumber \\
&& \left . +\frac{i}{M}\,\epsilon^{\mu\nu\alpha\beta}\,q_{\alpha}
s_{\beta}\, S_1(\nu,\,Q^2) \right . \nonumber \\
&& \left . + \frac{i}{M^3}\,\epsilon^{\mu\nu\alpha\beta}\,q_{\alpha} (p\cdot q\
s_{\beta}-s\cdot q\ p_{\beta})\, S_2(\nu,\,Q^2) \right \}\, ,
\end{eqnarray}
where $\epsilon_{0123} = +1$, $p$ is the nucleon momentum, and $s$ is the
nucleon covariant spin vector that satisfies $s \cdot p = 0$ and $s^2 = -1$. In
the following we are concerned only with the spin-dependent structure functions
of Eqs.~(\ref{DDeq2.3.19}) and (\ref{DDeq2.3.48}), which are related by
\begin{eqnarray}
\label{DDeq2.3.51} S_1(\nu,\,Q^2) & = & \frac{\nu\,M}{\nu^2+Q^2}\left(g_{TT}(\nu,\,Q^2) +
\frac{Q}{\nu}\,g_{LT}(\nu,\,Q^2)\right) \, , \nonumber \\
S_2(\nu,\,Q^2) & = & - \frac{M^2}{\nu^2+Q^2}\left(g_{TT}(\nu,\,Q^2) -
\frac{\nu}{Q}\,g_{LT}(\nu,\,Q^2)\right) \, .
\end{eqnarray}

As in the case of real Compton scattering, the VVCS amplitude has to be symmetric under
the crossing operation, i.e., under the replacements
$\varepsilon_{\mu}'^{\ast}\leftrightarrow\varepsilon_{\mu}$,
$q_{\mu}\rightarrow-q_{\mu}$, and in particular for $\nu=p\cdot q/M\rightarrow -\nu$. An
inspection of Eqs.~(\ref{DDeq2.3.19}) and (\ref{DDeq2.3.48}) therefore yields the results
\begin{equation}
\begin{array}{lllllll}
S_1(\nu,Q^2) & = & S_1(-\nu,Q^2) \ ,&
 S_2(\nu,Q^2) & = & - S_2(-\nu,Q^2)\ , \\
g_{TT}(\nu,Q^2) & = & -g_{TT}(-\nu,Q^2) \ ,& g_{LT}(\nu,Q^2) & = & g_{LT}(-\nu,Q^2) \ .
\end{array}
\end{equation}

\subsection*{4.2\quad Dispersion Relations and Sum Rules}

In order to set up dispersion relations, we have to construct the imaginary parts of the
amplitudes with contributions from both elastic and inelastic scattering. The elastic
contributions are obtained from the direct and crossed Born diagrams with the usual form
of the electromagnetic vertex for the transition $\gamma^* (q) + N(p) \to N(p + q)$,
\beqn \Gamma^\mu \;=\; F_D(Q^2) \, \gamma^\mu \;+\; F_P(Q^2) \, i \sigma^{\mu
\nu} \frac{q_\nu}{2 M} \; ,
\label{eq:bornvvcs}
\eeqn
with $F_D$ and $F_P$ denoting the nucleon Dirac and Pauli form factors,
respectively. The straightforward calculation yields the result~\cite{DPV03}
\begin{eqnarray}
\label{DDeq2.3.21} g_{TT}^{\mbox{\sc{Born}}} (\nu,\,Q^2) & = &
-\frac{\alpha_{fs}\nu}{2M^2} \left (F_P^2 + \frac{Q^2}{\nu^2-\nu_B^2+i \varepsilon}
\,G_M^2\right ) \, ,
\nonumber \\
g_{LT}^{\mbox{\sc{Born}}} (\nu,\,Q^2) & = & \frac{\alpha_{fs}Q}{2M^2} \left (F_DF_P -
\frac{Q^2}{\nu^2-\nu_B^2+i \varepsilon} \,G_EG_M \right ) \, ,
\end{eqnarray}
where $\nu_B=Q^2/2M$. We have split the elastic amplitudes of Eq.~(\ref{DDeq2.3.21}) into
a real contribution and a complex one containing the electric and magnetic Sachs form
factors $G_E\,(Q^2) = F_D\,(Q^2)-\tau\,F_P(Q^2)\ ,
 \quad G_M\,(Q^2) = F_D\,(Q^2)+F_P(Q^2)$,
with $\tau=Q^2/4M^2$. These form factors are normalized to $G_E(0)=e_N\ ,\quad
G_M(0)=e_N+\kappa_N$, where $e_N$ and $\kappa_N$ are the charge (in units of
$e$) and the a.m.m. of the respective nucleon. We also note that perturbative
QCD predicts $F_D(Q^2)\sim Q^{-4}$ and $F_P(Q^2)\sim Q^{-6}$ if
$Q^2\rightarrow\infty$.

Equation~(\ref{DDeq2.3.21}) shows us that the transition from real to virtual
photons is not straightforward, because the limits $Q^2\rightarrow 0$ and
$\nu\rightarrow 0$ cannot be interchanged~\cite{Ji93}. The Born amplitudes of
VVCS have poles at $\nu=\pm\nu_B\mp i\varepsilon$, corresponding to s- and
u-channel elastic scattering. In particular, the amplitudes are complex at this
pole, because the virtual photon can be absorbed at $\nu=\nu_B=Q^2/2M$ or
$x=1$.

In more physical terms, the crucial difference between a (space-like) virtual
and a real photon is the fact that the former can be absorbed by a charged
particle whereas the latter can only be ``absorbed'' at zero frequency,
$\nu=0$. At this point the real photon amplitudes can be expanded in a Taylor
series whose leading terms are determined by the Born terms; in particular,
$g_{TT}^{\rm{Born}}(\nu,0)$ reproduces exactly the leading term of the
spin-flip amplitude $g(\nu)$ of Eq.~(\ref{DDeq2.2.13}). The GDH sum rule is
then obtained by equating that series to an expansion of the dispersion
integral that involves imaginary parts from the inelastic processes for
$\nu>\nu_0$. The case of a virtual photon differs in two aspects: The imaginary
parts stem from both inelastic and elastic processes, and the real parts of the
amplitudes have two poles in the $\nu$ plane.

The inelastic contributions are regular functions in the complex $\nu$-plane
except for cuts from $-\infty$ to $-\nu_0$ and $+\nu_0$ to $+\infty$. The
optical theorem relates the inelastic contributions to the partial cross
sections of inclusive electroproduction,
\begin{eqnarray}
\label{DDeq2.3.20} {\mbox{Im}}\ g_{TT}(\nu,\,Q^2) = \frac{K}{4\pi}\sigma_{TT}(\nu,\,Q^2)
\ , \qquad {\mbox{Im}}\ g_{LT}(\nu,\,Q^2) = \frac{K}{4\pi}\sigma_{LT}(\nu,\,Q^2) \ ,
\end{eqnarray}
for energies above pion threshold, $\nu>\nu_0=m_{\pi}+(m_{\pi}^2+Q^2)/2M$. We
note that the products $K\,\sigma_{TT/LT}$ are independent of the choice of
$K$, because they are directly proportional to the measured cross section (see
Eqs.~(\ref{DDeq2.3.6})$-$(\ref{DDeq2.3.8})). Of course, the natural choice at
this point would be $K=K_G=|{\bq}|$, because we expect the photon
three-momentum on the RHS of Eq.~(\ref{DDeq2.3.20}). However, for reasons
explained above, we have chosen $K=K_H$.

The imaginary parts of the relativistic spin amplitudes follow from
Eqs.~(\ref{DDeq2.3.51}) and (\ref{DDeq2.3.20}),
\begin{eqnarray}
\label{DDeq2.3.55} {\rm Im}\ S_1 & = & \frac{\nu\, M}{\nu^2+Q^2}\,\frac{K}{4\pi}\,
\left(\sigma_{TT}+\frac{Q}{\nu}\,\sigma_{LT}\right )= \frac{e^2}{4M}\,\frac{M}{\nu}\,g_1
\ ,
\nonumber \\
{\rm Im}\ S_2 & = & -\frac{M^2}{\nu^2+Q^2}\,\frac{K}{4\pi}\,
\left(\sigma_{TT}-\frac{\nu}{Q}\,\sigma_{LT}\right )=
\frac{e^2}{4M}\,\frac{M^2}{\nu^2}\,g_2 \ .
\end{eqnarray}
As was pointed out in Ref.~\cite{DPV03}, this definition does not contain a
singularity at $\nu=\pm iQ$, because this point corresponds to the Siegert
limit, where $\sigma_{TT}=\pm i\sigma_{LT}$.

We next turn to the sum rules for the spin dependent VVCS amplitudes (see also
Ref.~\cite{JiO01} and references therein). Assuming an appropriate high-energy
behavior, the spin-flip amplitude $g_{TT}$ (which is odd in $\nu$) satisfies an
unsubtracted dispersion relation as in Eq.~(\ref{DDeq2.2.9}),
\beqn \label{DDeq2.3.28} {\mbox{Re}}\ g_{TT}\,(\nu,\,Q^2) = \frac{2\nu}{\pi}\,
{\mathcal{P}}\,\int_{0}^{\infty}\,\frac {{\mbox{Im}}\ g_{TT}\,(\nu',\,Q^2)}{\nu'^2-\nu^2}
\, d\nu'\ . \eeqn
Separating the contributions of the elastic scattering at $\nu'=\nu_B$ from the inelastic
contribution at $\nu'>\nu_0$ as given by Eq.~(\ref{DDeq2.3.20}), we obtain~:
\begin{eqnarray}
\label{DDeq2.3.29} {\mbox{Re}}\left[ \
g_{TT}\,(\nu,Q^2)-g_{TT}^{\mbox{\sc{pole}}}(\nu,Q^2)\right ]
 & = &
\frac{\nu}{2\pi^2}\,{\mathcal{P}}
\int_{\nu_0}^{\infty}\frac{K(\nu',Q^2)\,\sigma_{TT}(\nu',Q^2)}
{\nu'^2-\nu^2} \, d\nu' \\
& = & \frac{2 \, \alpha_{fs} }{M^2}  \, I_{TT}(Q^2) \; \nu + \gamma_{TT}(Q^2) \; \nu^3 +
{\mathcal{O}}(\nu^5) \, . \nonumber
\end{eqnarray}
Because the RHS of this equation is now a regular function, it can be expanded
in a Taylor series whose leading terms yield a generalization of the GDH sum
rule,
\beqn \label{DDeq2.3.30} I_{TT}(Q^2) &=& \frac{M^2}{\pi \, e^2}\, \int_{\nu_0}^{\infty}\,
\frac{K(\nu, \, Q^2)}{\nu} \,
\frac{\sigma_{TT}\,(\nu,\,Q^2)}{\nu}\,d\nu \, , \nonumber \\
&=& \frac{2 \, M^2}{Q^2}\, \int_{0}^{x_0}\,dx \, \left\{ g_1\,(x,\,Q^2) - \frac{4
M^2}{Q^2} \, x^2 \, g_2\,(x,\,Q^2) \right\} \, , \eeqn
and similarly a generalized form of the FSP,
\beqn \label{eq:gammao} \gamma_{TT}\,(Q^2) &=& \frac{1}{2\pi^2}\, \int_{\nu_0}^{\infty}\,
\frac{K(\nu, \, Q^2)}{\nu} \,
\frac{\sigma_{TT}\,(\nu,\,Q^2)}{\nu^3}\,d\nu \, , \nonumber \\
&=& \frac{e^2 \, 4M^2}{\pi \, Q^6}\,\int_{0}^{x_0}\,dx \, x^2 \, \left\{ g_1\,(x,\,Q^2)
-\frac{4 M^2}{Q^2} \, x^2 \, g_2\,(x,\,Q^2) \right\} \, . \eeqn
Comparing these integrals with the case of real photons, Eqs.~(\ref{DDeq2.2.15}) and
(\ref{DDeq2.2.16}), we find $I_{TT} (Q^2=0) = -\frac{1}{4}\kappa_N^2$ and $\gamma_{TT}
(Q^2=0) = \gamma_0$.

We next turn to the amplitude $g_{LT}(\nu, Q^2)$, which is even in $\nu$. Assuming that
$g_{LT}$ can be described by an unsubtracted dispersion relation, a similar procedure
yields the integral
\beqn \label{eq:i3int} I_{LT}(Q^2) & = & \frac{M^2}{\pi \,e^2}\,
\int_{\nu_0}^{\infty}\,\frac{K(\nu \, , Q^2)}{\nu} \,
\frac{1}{Q} \, \sigma_{LT}\,(\nu,\,Q^2)\,d\nu\  \nonumber\\
&=& \frac{2 \, M^2}{Q^2}\,\int_{0}^{x_0}\,dx \, \left\{ g_1\,(x,\,Q^2) \,+\,
g_2\,(x,\,Q^2) \right\} \, , \eeqn
and a generalized longitudinal-transverse polarizability
\beqn \label{eq:deltalt} \delta_{LT}\,(Q^2) &\;=\;& \frac{1}{2\pi^2}\,
\int_{\nu_0}^{\infty}\,\frac{K(\nu, \, Q^2)}{\nu} \,
\frac{\sigma_{LT}(\nu\, ,Q^2)}{Q\,\nu^2}\,d\nu  \nonumber \\
&\;=\;& \frac{e^2 \, 4M^2}{\pi \, Q^6}\,\int_{0}^{x_0}\,dx \, x^2 \, \left\{
g_1\,(x,\,Q^2) \,+\, g_2\,(x,\,Q^2) \right\} \, . \eeqn
Both functions are finite in the real photon limit, because $\sigma_{LT}/Q$ is
finite for $Q^2\rightarrow 0$. In the limit of large $Q^2$,
Wandzura~\&~Wilczek~\cite{WW77} have shown that $g_1+g_2$ can be expressed in
terms of the twist-2 spin structure function $g_1$ if the dynamical (twist-3)
quark-gluon correlations are neglected,
\beqn \label{eq:ww} g_1\,(x, Q^2) \,+\, g_2\,(x, Q^2) \;=\; \int_{x}^{1}\, dy \;
\frac{g_{1}\,(y,\,Q^2)}{y} \, . \eeqn

By use of Eq.~(\ref{DDeq2.3.55}) we next construct the dispersion relations for the spin
dependent amplitudes $S_1$ and $S_2$. The amplitude $S_1$ is even in $\nu$, and an
unsubtracted dispersion relation reads
\begin{eqnarray}
&&\hspace{-.25cm}{\mbox{Re}}\ S_1(\nu,\,Q^2) -
{\mbox{Re}}\ S_1^{\mbox{\sc{pole}}}(\nu,\,Q^2) =  \\
&&\hspace{-.25cm} \frac{2\alpha_{fs} }{M} I_1(Q^2) \hspace{-.1cm}+ \hspace{-.1cm}
\frac{2\alpha_{fs} }{M\,Q^2} \left[ I_{TT}(Q^2) - I_1(Q^2) + \frac{M^2Q^2}{2\alpha_{fs}}\
\delta_{LT}(Q^2) \right]\hspace{-.1cm} \nu^2  + {\mathcal{O}}(\nu^4) \, ,\nonumber
\label{eq:s1lex}
\end{eqnarray}
with
\beqn I_1(Q^2) &\equiv& \frac{2M^2}{Q^2}\int_0^{x_0}
g_1(x,\,Q^2)\,dx \nonumber \\
&=& \frac{M^2}{\pi \,e^2}\, \int_{\nu_0}^{\infty}\,\frac{K(\nu, Q^2)}{(\nu^2 + Q^2)}
\left\{\sigma_{TT}\,(\nu,\,Q^2) \,+\, \frac{Q}{\nu} \, \sigma_{LT}\,(\nu,\,Q^2) \right\}
\,d\nu \, . \label{eq:I1} \eeqn

The second spin-dependent VVCS amplitude $S_2$ is odd in $\nu$. If we further
assume that the behavior of $S_2$ for $\nu \to \infty $ is given by $S_2(\nu,
Q^2) \to \nu^{\alpha_2}$ with $\alpha_2 < -1$, there should also exist an
unsubtracted dispersion relation for the even function $\nu \, S_2$. If we now
subtract the dispersion relations for $S_2$ and $\nu S_2$, we obtain a
``superconvergence relation'' that is valid for {\it any} value of $Q^2$,
\beqn \label{eq:drbcs2} \int_0^{\infty} {\rm Im}\ S_2(\nu,\,Q^2)\,d\nu = 0 \quad
{\mbox{or}} \quad \int_{0}^{1}\, dx \, g_{2}\,(x,\,Q^2)=0 \, , \eeqn
i.e., the sum of all elastic and inelastic contributions vanishes.

Equation~(\ref{eq:drbcs2}) is known as the Burkhardt-Cottingham (BC) sum
rule~\cite{BC70}. The BC sum rule was shown to be fulfilled in the case of QED
to lowest order in $\alpha_{fs}$~\cite{Tsai75}. It is also fulfilled in
perturbative QCD when calculated for a quark target to first order in
$\alpha_s$~\cite{Alt94}. The BC sum rule follows from the Wandzura-Wilczek
relation, as can be easily proven by integrating Eq.~(\ref{eq:ww}) over all
values of the Bjorken parameter $x$.

Separating the elastic and inelastic contributions in Eq.~(\ref{eq:drbcs2}) and
using Eqs.~(\ref{DDeq2.3.21}) - (\ref{DDeq2.3.55}), we may cast the BC sum rule
in the form
\beqn \label{DDeq2.3.46} I_2(Q^2) \;\equiv\; \frac{2M^2}{Q^2}\int_0^{x_0}g_2(x,\,Q^2)\,dx
\;=\; \frac{1}{4} \, F_P(Q^2) \, \left( F_D(Q^2) + F_P(Q^2) \right) \, . \eeqn
Alternatively this sum rule can be written in terms of the Sachs form factors and the
absorption cross sections,
\beqn \label{eq:drbcg4} I_2(Q^2)&\;=\;&\frac{M^2}{\pi \, e^2 }\,
\int_{\nu_0}^{\infty}\,\frac{K(\nu,\,Q^2)}{\nu^2 + Q^2} \, \left\{ \,-
\sigma_{TT}(\nu,\,Q^2) \,+\, \frac{\nu}{Q} \, \sigma_{LT}(\nu,\,Q^2) \, \right\} \, d \nu
\nonumber \\
&\;=\;& \frac{1}{4}\ \frac{G_M(Q^2)(G_M(Q^2)-G_E(Q^2))}{1+\tau}\, . \eeqn

The low energy expansion of the dispersion relation for $(\nu S_2) - (\nu
S_2)^{{\mbox{\sc{pole}}}}$ takes the form
\begin{eqnarray}
&&{\mbox{Re}}\ \left(\nu \, S_2(\nu,\,Q^2)\right) -
{\mbox{Re}}\ (\nu \, S_2(\nu,\,Q^2))^{\mbox{\sc{pole}}} = \nonumber \\
&& 2 \alpha_{fs} \, I_2(Q^2) - \frac{2 \alpha_{fs}}{Q^2} \left( I_{TT}(Q^2) - I_1(Q^2)
\right) \, \nu^2
 \\
&& + \frac{2\alpha_{fs}}{Q^4} \left[
 I_{TT}(Q^2) \hspace{-.05cm} - \hspace{-.05cm} I_1(Q^2)
\hspace{-.05cm} + \hspace{-.05cm} \frac{M^2Q^2}{2\alpha_{fs}} \left( \delta_{LT}(Q^2)
\hspace{-.05cm} - \hspace{-.05cm} \gamma_{TT}(Q^2) \right) \right] \hspace{-.1cm}\nu^4
 \hspace{-.05cm} + \hspace{-.05cm} {\mathcal{O}}(\nu^6)\ , \nonumber
\label{eq:s2lex}
\end{eqnarray}
in terms of the integrals and spin polarizabilities introduced above. We note
that the relation $I_{TT}'(0) - I_1'(0) = M^2 / (2 \, \alpha_{fs}) \, \cdot
\left( \gamma_{TT}(0) - \delta_{LT}(0) \right)$ ensures that the $\nu^4$ term
in $\nu S_2$ has no singularity at $Q^2 = 0$ (see Ref.~\cite{Dre01}).

In concluding this section we point out that the values of all the discussed integrals at
the real photon point are determined by the charge and the a.m.m.,
\be \label{I_LT_0}
\begin{array}{lllll}
I_1(0)& = & -\frac{1}{4}\kappa_N^2\ ,\quad
I_2(0)& = & \frac{1}{4}\kappa_N (e_N + \kappa_N)\ , \\
I_{TT}(0) & = & -\frac{1}{4}\kappa_N^2\ , \quad I_{LT} (0) & = & \frac{1}{4} e_N \kappa_N
\ .
\end{array}
\ee
%

\subsection*{4.3\quad The Helicity Structure of the Cross Sections}

Let us now discuss the helicity-dependent cross sections $\sigma_{TT}$ and
$\sigma_{LT}$ which determine the various integrals of the previous section.
For momentum transfer $Q^2\lesssim0.5$~GeV$^2$, the bulk contribution to these
cross sections is due to the production of a single pion whose multipole
content is reasonably well known in the resonance region $W\lesssim 2$~GeV. The
threshold region is dominated by s-wave production ($E_{0+},\ S_{0+}$)
accompanied by much smaller contributions of p waves ($M_{1\pm},\ E_{1\pm},\
S_{1\pm}$). Low-energy theorems, the predictions of ChPT, and several new
precision experiments have provided a solid basis for the multipole
decomposition in that region. The data basis is also quite reliable in the
first resonance region. Although the leading $M_{1+}$ multipole drops with
$Q^2$ somewhat faster than the dipole form factor, it dominates that region up
to large momentum transfers. The ratio $R_E = E_{1+}/M_{1+}$ remains in the few
percent range up to $Q^2\approx 4$~GeV$^2$~\cite{Fro99}, far from the
prediction of perturbative QCD, $R_E\rightarrow 1$~\cite{Bro81}. On the other
hand, the ratio $R_S = S_{1+}/M_{1+}$ increases in absolute value, from
$R_S\approx-5\%$ for small $Q^2$ to $R_S\approx-15\%$~\cite{Fro99} at
$Q^2\approx 4$~GeV$^2$. Unfortunately, the multipole decomposition in the
higher resonance regions is still not well under control. In particular, there
is as yet little reliable information on $\sigma_{LT}$, except that this cross
section has been found to be small. This fact is not really consoling in the
context of the sum rules, because the integral $I_{LT}$ (see
Eq.~(\ref{eq:i3int})) does not converge well. However, great improvements in
the data basis are expected from the wealth of ongoing and planned polarization
experiments.

\begin{figure}
\centerline{\psfig{figure=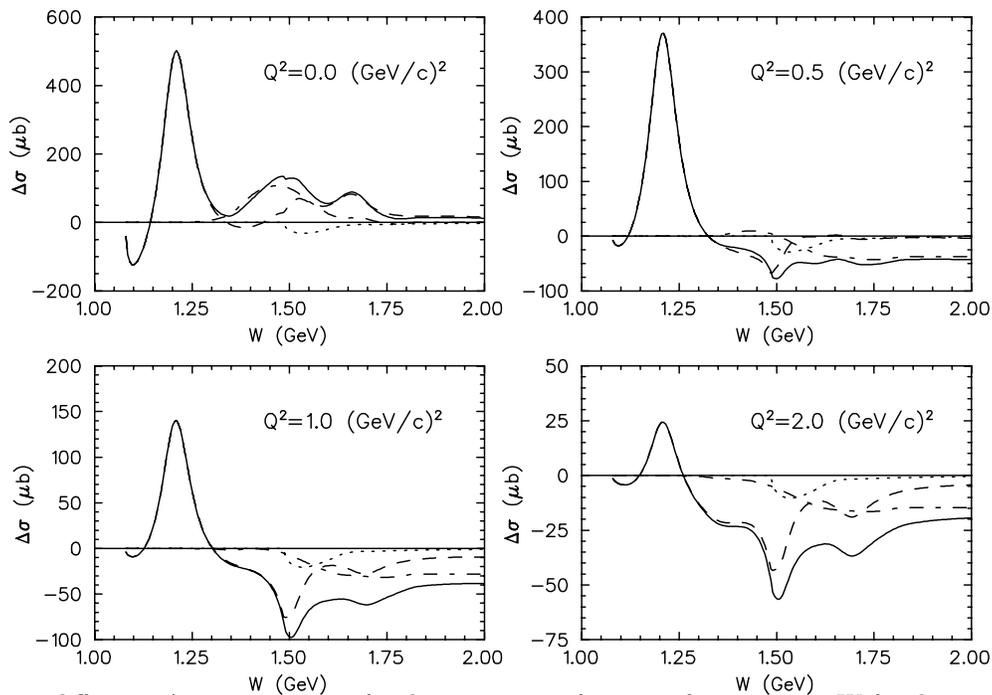,width=13cm,angle=90}} \vspace{-.5cm}
\caption{The helicity difference $\Delta\sigma =\sigma_{3/2}-\sigma_{1/2}$ for
the proton as a function of c.m. energy $W$ for the momentum transfers
$Q^2=0,0.5,1.0$, and 2.0~GeV$^2$ according to Ref.~\cite{Maid03}. The figure
shows the total helicity difference (solid line) and the contributions of the
channels with one-pion (dashed line), more-pion (dash-dotted line), and $\eta$
(dotted line) production.}
\label{Delta_sigma}
\end{figure}

The one-pion contribution to $\Delta\sigma$ of the proton is displayed in
Fig.~\ref{Delta_sigma} as a function of the c.m. energy $W$ for different
momentum transfers. The figure shows negative values near threshold, because
there the pions are essentially produced in s waves leading to states of
helicity $1/2$ and $\sigma_{1/2}$ dominance. This s-wave production is strongly
suppressed, however, with increasing values of $Q^2$. The $\Delta(1232)$ yields
large positive values because of the strong M1 transition, which aligns the
quark spins (paramagnetism). In the case of real photons ($Q^2=0$), the
helicity difference in the second and third resonance regions carries the same
sign as in the first resonance region. However, $\Delta\sigma$ changes sign at
some value below $Q^2=0.5$~GeV$^2$, and becomes negative and large relative to
its value in the first resonance region. Let us study this effect in more
detail for the N*(1520) with multipoles $E_{2-}$ and $M_{2-}$ ($J^P
=\frac{3}{2}^-$). According to Eq.~(\ref{sigmaTT}) we find $\sigma_{3/2} -
\sigma_{1/2} \sim |E_{2-}|^2 + 6\,{\mbox{Re}}\,(E_2^{\ast}M_{2-}) -
3|M_{2-}|^2$. This value is positive at the real photon point where the
electric dipole radiation (E1) dominates over the magnetic quadrupole radiation
(M2). Yet as the magnetic term increases with $Q^2$, the helicity difference
runs through a zero at $Q^2\approx 0.4$~GeV$^2$, and eventually the cross
section $\sigma_{1/2}$ dominates. The latter finding is a prediction of
perturbative QCD since only states with helicity $1/2$ can be produced by
absorption on an individual free quark, and it is also corroborated by the
analysis of various experiments~\cite{Fro99,Joo02,Tia03}.

Figure~\ref{Delta_sigma} also shows an overall decrease of the cross sections
with increasing momentum transfer $Q^2$. The reason is, of course, that the
coherent resonance effects are of long range and are therefore damped by form
factors. It is also evident that the upper part of the spectrum becomes more
and more important. Finally, if $Q^2$ reaches values beyond 4~GeV$^2$, the
resonance structures become small fluctuations on top of the low-energy tail of
DIS.

\subsection*{4.4\quad Recent Data for GDH-like Integrals}

The integrals $I_1$ of Eq.~(\ref{eq:I1}) and $I_2$ of Eq.~(\ref{DDeq2.3.46}) can be
expressed in terms of the {\it inelastic} contributions to the first moments $\Gamma_1$
and $\Gamma_2$ of the spin structure functions,
\beqn \label{eq:Aqhighq} I_i(Q^2) =  \frac{2 \, M^2}{Q^2} \int_{0}^{x_0}\,dx \,
g_i\,(x,\,Q^2) =
 \frac{2 \, M^2}{Q^2} \; \Gamma_i(Q^2) \ .
\label{Gammai} \eeqn

The definition of $\Gamma_i(Q^2)$ as the first moment of $g_i$ includes, of
course, both elastic and inelastic contributions. However, the elastic
contribution is generally well known at small $Q^2$, and at large $Q^2$ it
vanishes like $Q^{-10}$ (see, e.g., the RHS of Eq.~(\ref{DDeq2.3.46})). In the
following discussion of asymptotia we may therefore safely neglect the elastic
contribution to these observables. The operator product expansion (OPE)
predicts~\cite{Ji93} the following asymptotic form for the first moments of
$g_1$ and $g_2$:
\begin{equation}
\label{Gamma1Q2} \Gamma_1(Q^2) = \tilde{\Gamma}_1 (Q^2) +
{\mathcal{O}}\,\left(\frac{M^2}{Q^2}\right)\ ,\quad \Gamma_2(Q^2) =
{\mathcal{O}}\,\left(\frac{M^4}{Q^4}\right)\ ,
\end{equation}
where $\tilde{\Gamma}_1$ is the asymptotic (twist-2) contribution with only a logarithmic
dependence on $Q^2$. For large $Q^2$ the isovector combination fulfills the Bjorken sum
rule~\cite{Bjo66},
\beqn \label{eq:bjsr2} \tilde{\Gamma}_1^p (Q^2) - \tilde{\Gamma}_1^n (Q^2) = \frac{1}{6}
\, g_A \left\{ 1 -  \frac{\alpha_s(Q^2)}{\pi} + {\mathcal{O}}\,(\alpha_s^2)\right\} \ ,
\eeqn
where $g_A$ is the axial-vector coupling constant and $\alpha_s\,(Q^2)$ is the
running coupling constant of the strong interaction. The Bjorken sum rule was
originally derived from current algebra, and therefore is also a prediction of
QCD. Although the experimental data for $\tilde{\Gamma}_1^p$ and
$\tilde{\Gamma}_1^n$ differ from the originally expected values, the Bjorken
sum rule is well established. Evaluating Eq.~(\ref{eq:bjsr2}) at
$Q^2=5$~GeV$^2$, using three light flavors in $\alpha_s$, and fixing this
constant at the mass of the $Z$ boson, one obtains~\cite{Lar91}
$\tilde{\Gamma}_1^p-\tilde{\Gamma}_1^n=0.182\pm0.005$, while a fit to all the
available DIS data~\cite{Ant00} yields $\tilde{\Gamma}_1^p = 0.118 \pm 0.004
\pm 0.007$, $\tilde{\Gamma}_1^n = -0.058 \pm 0.005 \pm 0.008$, and hence
$\tilde{\Gamma}_1^p - \tilde{\Gamma}_1^n = 0.176 \pm 0.003 \pm 0.007$, in
agreement with the sum rule. The small value of $\tilde{\Gamma}_1^p$, on the
other hand, contradicts a simple quark model interpretation. This discrepancy
led to the ``spin crisis'' of the 1980s and taught us that less than half of
the nucleon's spin is carried by the quarks. With regard to the second spin
structure function we note that $\Gamma_2$ vanishes identically for all $Q^2$
if the BC sum rule holds. In particular the inelastic contribution would be
${\mathcal{O}}\,(M^{10}/Q^{10})$
 in the scaling limit as can be seen from Eq.~(\ref{DDeq2.3.46})
 and the discussion after Eq.~(\ref{DDeq2.3.21}).

Although at first glance the introduction of $I_i(Q^2)$ and $\Gamma_i(Q^2)$ for
the same physics content may look like a nuisance, both definitions serve a
purpose. The functions $\Gamma_i(Q^2)$ are appropriate to derive the higher
twist terms of the OPE (Eq.~(\ref{Gamma1Q2})), but at the same time they
suppress the visibility of resonance effects and completely hide the efforts of
the real photon work. Essentially the opposite is true for the functions
$I_i(Q^2)$, and since we want to highlight the resonance effects, we will
generally discuss these latter functions here.

\begin{figure}
\centerline{\psfig{figure=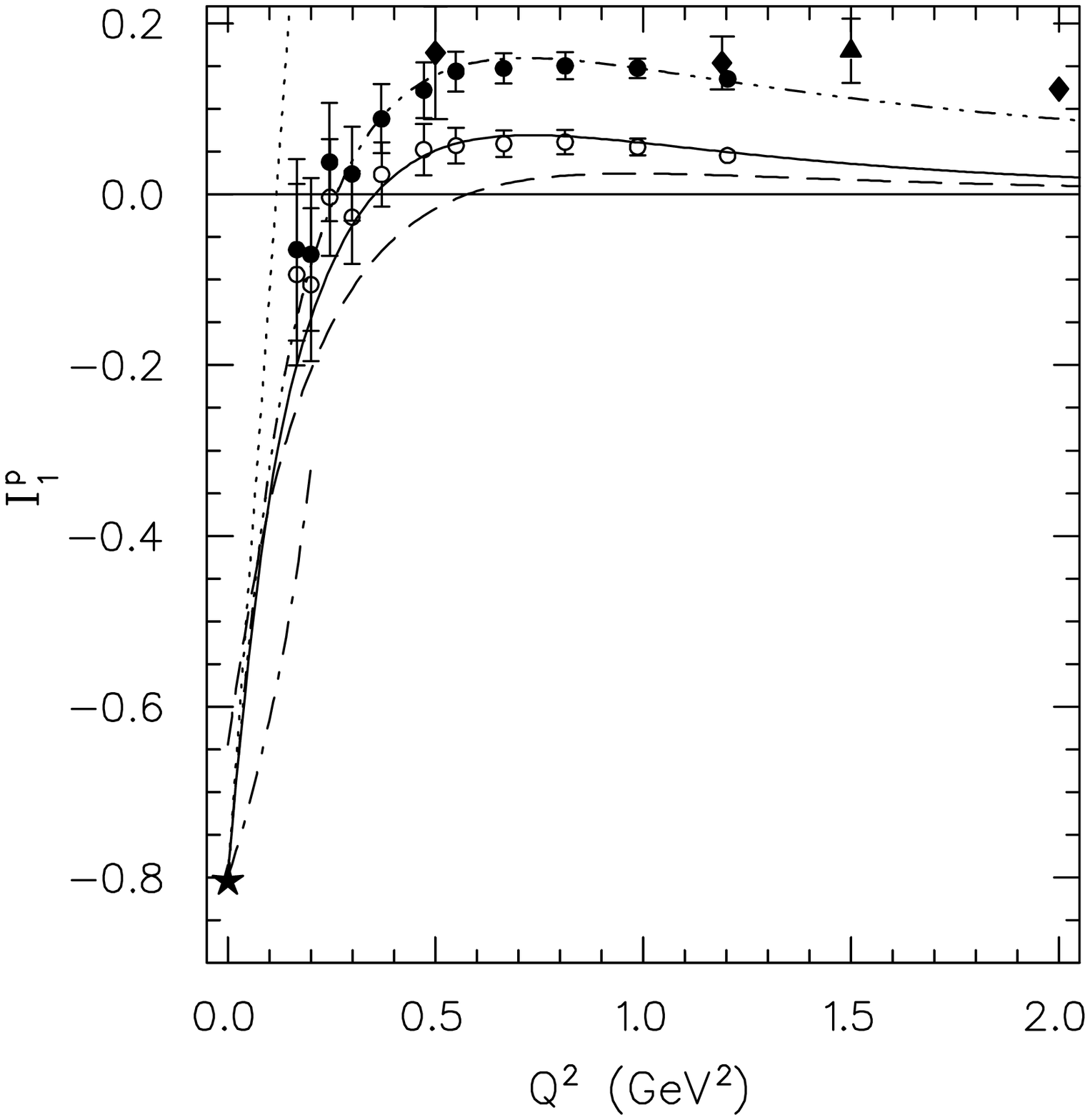,width=6.75cm}
\psfig{figure=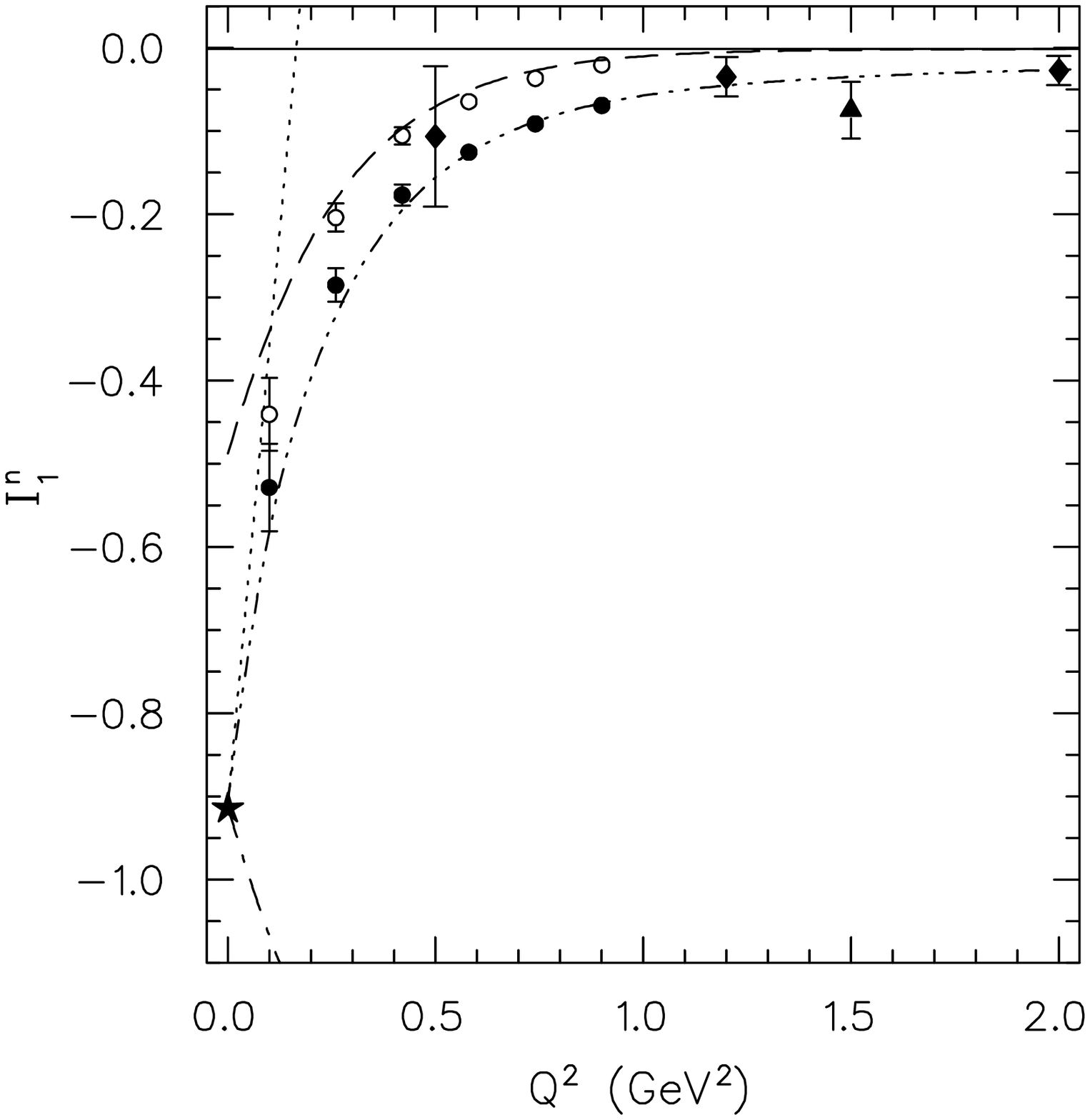,width=6.75cm}} \caption{ The $Q^2$ dependence of the
integrals $I_1$ as defined in Eq.~(\ref{eq:I1}). The open circles correspond to
the measured resonance contribution ($W<2$~GeV), the solid symbols include the
measured or estimated DIS contributions. The proton data (left) are obtained by
the CLAS Collaboration (circles, Ref.~\cite{Fat03}), SLAC (diamonds,
Ref.~\cite{Abe97}), and HERMES (triangles, Ref.~\cite{Air02}). The neutron data
(right) are from the JLab E94-010 Collaboration (circles, Ref.~\cite{Ama02}).
The result of MAID~\cite{Maid03} for the one-pion channel is given by the
dashed lines, the full line for the proton includes all channels integrated up
to $W=2$~GeV. Also shown are ${\mathcal{O}}(p^4)$ predictions of heavy baryon
ChPT (dotted lines, Ref.~\cite{JiK00a}) and relativistic baryon ChPT
(dash-dotted lines, Ref.~\cite{Ber02}). The dash-dot-dotted lines represent the
interpolating formula of Eq.~(\ref{eq:bi1}), the asterisks show the sum rule
values at $Q^2=0$ according to Eq.~(\ref{I_LT_0}). } \label{fig:i1p}
\end{figure}

Figure~\ref{fig:i1p} shows the $Q^2$ dependence of $I_1^p$ and $I_1^n$. The
rapid increase of $I_1^p$ from its large negative value
$-\frac{1}{4}\kappa^2_p=-0.80$ at $Q^2=0$ to positive values in the DIS region
is particularly striking. The recent data from JLab CLAS~\cite{Fat03} cover the
range $W\lesssim2$~GeV and 0.15~GeV$^2\lesssim Q^2\lesssim 1.5$~GeV$^2$ and
they clearly confirm a sign change of the integral at relatively small momentum
transfer. These data are in good agreement with the resonance estimate of MAID.
However, the integral up to $W=2$~GeV has not yet converged, and with
increasing values of $Q^2$ the DIS contributions above this energy become more
and more important. Altogether the figure shows a rather dramatic transition
from a resonance-dominated coherent photoabsorption at low $Q^2$ to an
incoherent partonic description at large $Q^2$. As we discussed in the real
photon case (Table~\ref{tab1}), the resonance region contributes $\sim95$~\% of
the GDH sum rule. However, these long-range coherent effects are strongly
damped by form factors and therefore decrease in importance relative to the
incoherent processes. As an example, the resonance contribution for the proton
has dropped to $\sim20$~\% already at $Q^2=2$~GeV$^2$, in agreement with the
analysis of Ref.~\cite{Ede00}.

The situation for the neutron is quite similar to the proton case, except that there is
no zero crossing. The MAID prediction for the one-pion channel is in good agreement with
the resonance data~\cite{Ama02} for $Q^2\gtrsim 0.3$~GeV$^2$. However it fails to
describe the measurements at the smaller values of $Q^2$, as is to be expected from the
discussion of the ``neutron puzzle'' in section 3.5.

In order to describe this gradual transition from coherent to incoherent
processes, Anselmino et al.~\cite{Ans89} proposed to parameterize the $Q^2$
dependence in the framework of vector meson dominance. This model was refined
by adding explicit resonance contributions~\cite{Bur92} and resulted in the
following phenomenological parameterization~:
\begin{equation}
I_1^{p, n}(Q^2) =  I_{1, res}^{p, n}(Q^2) + 2 M^2 \, \tilde{\Gamma}_{1}^{p, n} \left[ {1
\over (Q^2 + \mu^2)} - {{c^{p,n} \, \mu^2} \over (Q^2 + \mu^2)^2} \right] \, ,
\label{eq:bi1}
\end{equation}
where $I_{1, res}^{p, n}(Q^2)$ are the resonance contributions and
$\tilde{\Gamma}_{1}^{p, n}$ the asymptotic values for the first moments of
$g_1$. Furthermore the scale $\mu$ is assumed to be the vector meson mass, $\mu
= m_\rho$, and the parameters $c^{p,n}$ are chosen to reproduce the GDH sum
rule at $Q^2=0$,
\begin{equation}
c^{p, n} = 1 + {{\mu^2} \over {2 M^2}} \, {1 \over \tilde{\Gamma}_{1}^{p, n}} \left[
{{\kappa^2_{p,n} \over 4} + I_{1, res}^{p, n}(0)} \right] \, . \label{eq:bi2}
\end{equation}

We use the parameterization of Eq.~(\ref{eq:bi1}) with the recent experimental value for
$\tilde{\Gamma}_{1}$~\cite{Ant00}, and the resonance contribution follows from the
experimental GDH integral up to $W=2$~GeV, e.g., $I_{1, res}^{p}(0) = -0.95$, modified by
a $Q^2$ dependence as given by MAID. This parameterization is shown by the
dash-dot-dotted lines in Fig.~\ref{fig:i1p}, which give a rather good description of the
data and in particular of the sign change at $Q^2 \simeq 0.25$~GeV$^2$ for the proton.

In Fig.~\ref{fig:i1p} we also present the predictions of heavy baryon ChPT to
${\mathcal O}(p^4)$ from Ref.~\cite{JiK00a}  and relativistic baryon ChPT to
${\mathcal O}(p^4)$ from Ref.~\cite{Ber02}. Comparing these calculations to the
data for the GDH integrals, we find that the chiral expansion may only be
applied in a very limited range of $Q^2 \lesssim 0.05$~GeV$^2$. This is partly
explained by the above discussion about the importance of the $\Delta(1232)$
resonance. Of course, this resonance is tacitly taken into account by inserting
the a.m.m. as a LEC at the real photon point. However, the rapid change of the
integrals as a function of $Q^2$ comes about by an interplay of s-wave pion and
resonance production. The form factor associated with the former process should
indeed be described by the spatial extension of the pion loops in ChPT.
However, the strong N$\Delta$ transition form factor and the resonance width
require a dynamical treatment of the $\Delta$ as well as higher order terms in
the perturbation series.

Because the $\Delta(1232)$ contribution and other isospin 3/2 resonances drop
out in the isovector combination of the integrals, ChPT should be able to
describe the difference $I_1^p - I_1^n$ over a larger $Q^2$ range~\cite{Bur01}.
The predicted $Q^2$ dependence for this observable is indeed much smoother than
for $I_1^p$ and $I_1^n$ individually and is in better agreement with the
existing data. This may open the possibility to bridge the gap between the low
and high $Q^2$ regimes, at least for this particular observable. A real
quantitative test of ChPT should soon be possible by combining the low $Q^2$
data of the Hall~A and B Collaborations at JLab.

\begin{figure}
\centerline{\psfig{figure=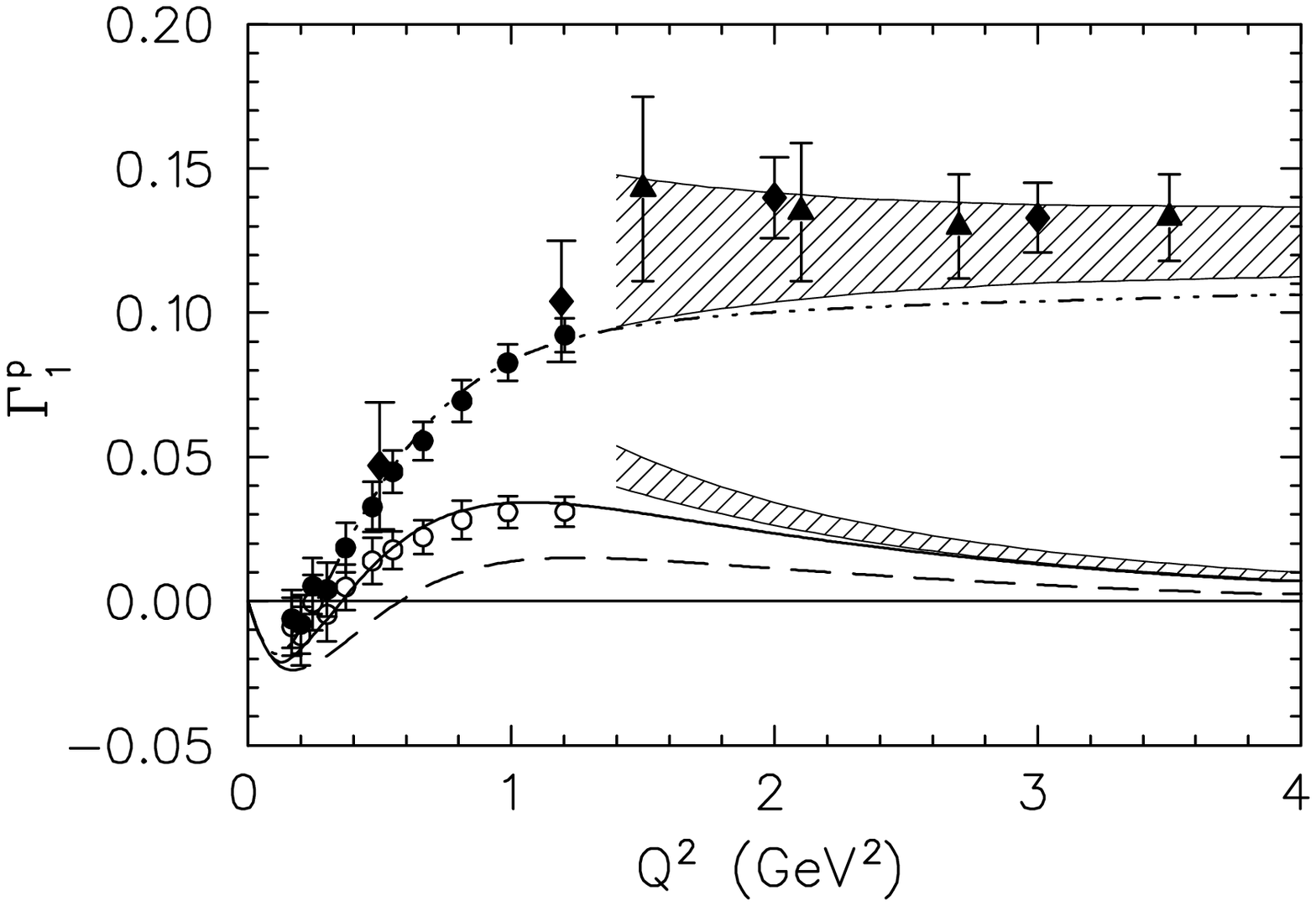,width=6.75cm}
\psfig{figure=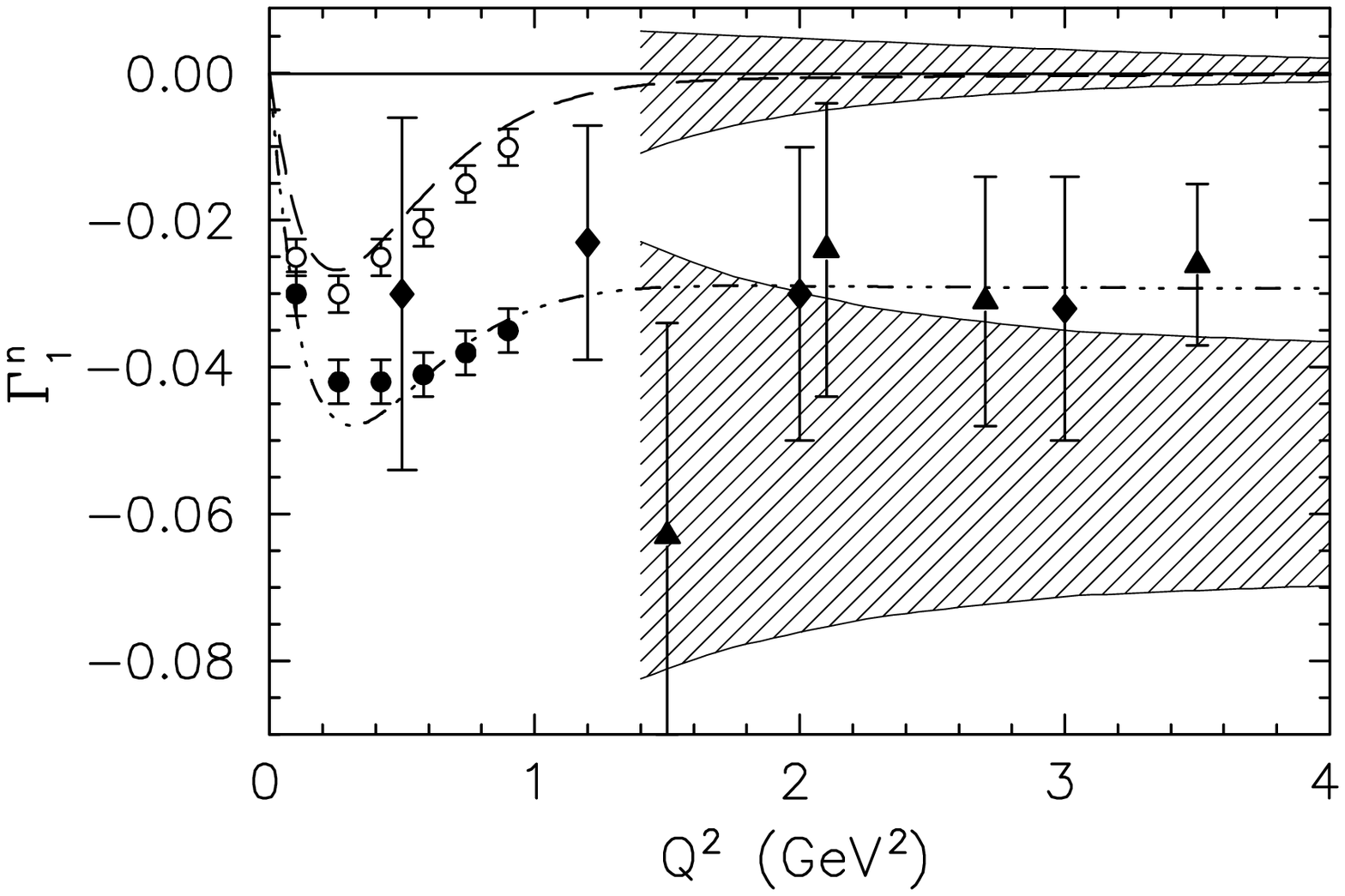,width=6.75cm}} \caption{The $Q^2$ dependence of the {\it
inelastic} contributions to the moments $\Gamma_1$ as defined in Eq.~(\ref{Gammai}). The
hatched bands represent an error estimate obtained by integrating the DIS structure
functions of Ref.~\cite{Blum02} over the full energy range (upper band) and over the
resonance region only (lower band). See Fig.~\ref{fig:i1p} for further notation. }
\label{fig:Gamma}
\end{figure}

The inelastic contributions to the moments $\Gamma_1^p$ and $\Gamma_1^n$ are
plotted in Fig.~\ref{fig:Gamma}. As has been stated, this representation is
more appropriate if one wants to study the transition to DIS. The upper shaded
band in the left panel represents the error estimate for the DIS structure
function of the proton. It is centered on the asymptotic value
$\tilde{\Gamma}_1^p\approx 0.12$. The lower shaded band is obtained by
integration of the DIS structure function only up to $W=2$~GeV, which matches
reasonably well the JLab data at $Q^2\approx 1$~GeV$^2$. Whereas the DIS
contribution is positive for the proton, it is negative in the case of the
neutron as shown in the right panel of Fig.~\ref{fig:Gamma}, and as a
consequence there is no zero crossing in the case of $\Gamma_1^n$. We also
observe that the resonance contribution to this moment decreases rapidly with
increasing $Q^2$, even faster than in the proton case.

\begin{figure}
\centerline{\psfig{figure=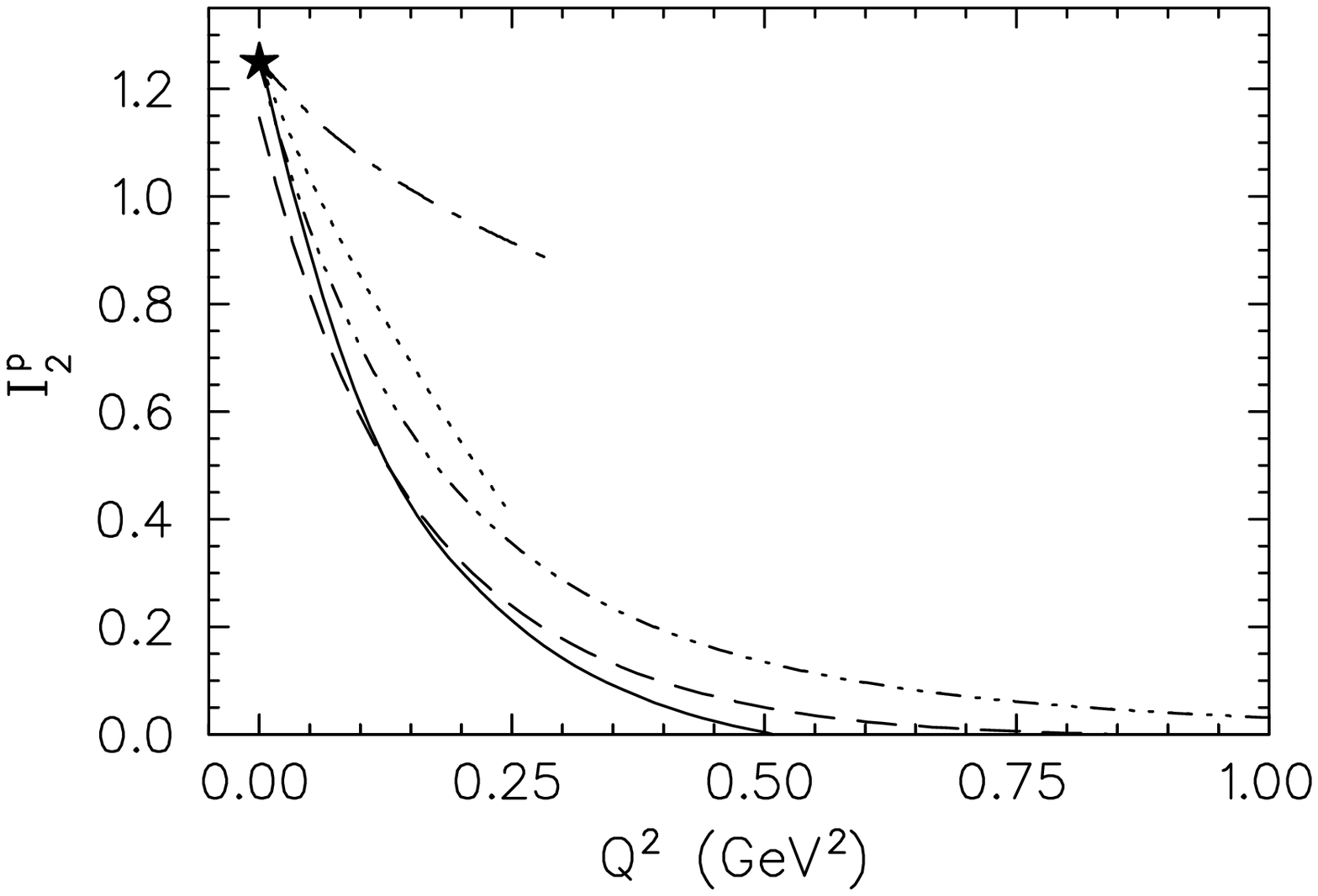,width=6.75cm}
\psfig{figure=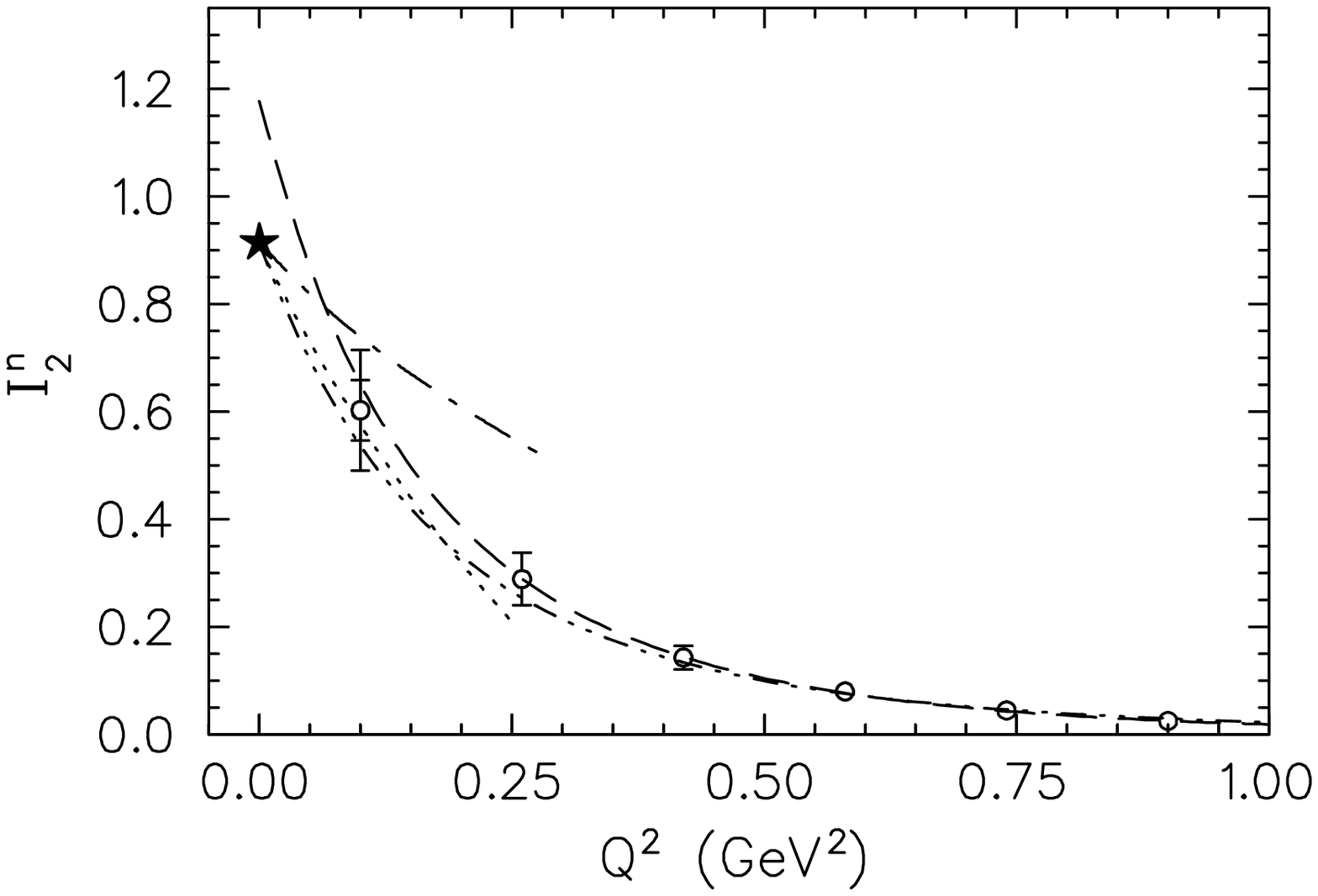,width=6.75cm}} \caption{The $Q^2$ dependence of the
integrals $I_2$ as defined in Eq.~(\ref{eq:drbcg4}). The neutron data were
obtained by the JLab E94-010 Collaboration~\cite{Ama03}. Dash-dot-dotted line:
Burkhardt-Cottingham sum rule of Eq.~(\ref{eq:drbcg4})). For further notation
see Fig.~\ref{fig:i1p}. } \label{fig:I2p}
\end{figure}

Figure~\ref{fig:I2p} compares the MAID prediction for $I_2(Q^2)$ with the BC
sum rule value and the recent neutron data of the JLab E94-010
Collaboration~\cite{Ama03}. In the case of the proton, the one-pion channel
nearly saturates the BC sum rule for small $Q^2$, but at intermediate values of
$Q^2$ the MAID calculation starts to fall short of the sum rule because of
higher continua and DIS contributions. On the other hand, for the neutron,  the
MAID prediction overshoots the BC sum rule at small $Q^2$ but agrees with the
data for $Q^2\gtrsim0.1$~GeV$^2$. At the much larger momentum transfer of
$Q^2=5$~GeV$^2$, the SLAC E155 Collaboration has recently evaluated the BC
integral in the measured region of $0.02\le x\le0.8$. The results indicate a
small deviation from the sum rule, but this could be well compensated by
contributions from the unmeasured region.

\begin{figure}
\centerline{\psfig{figure=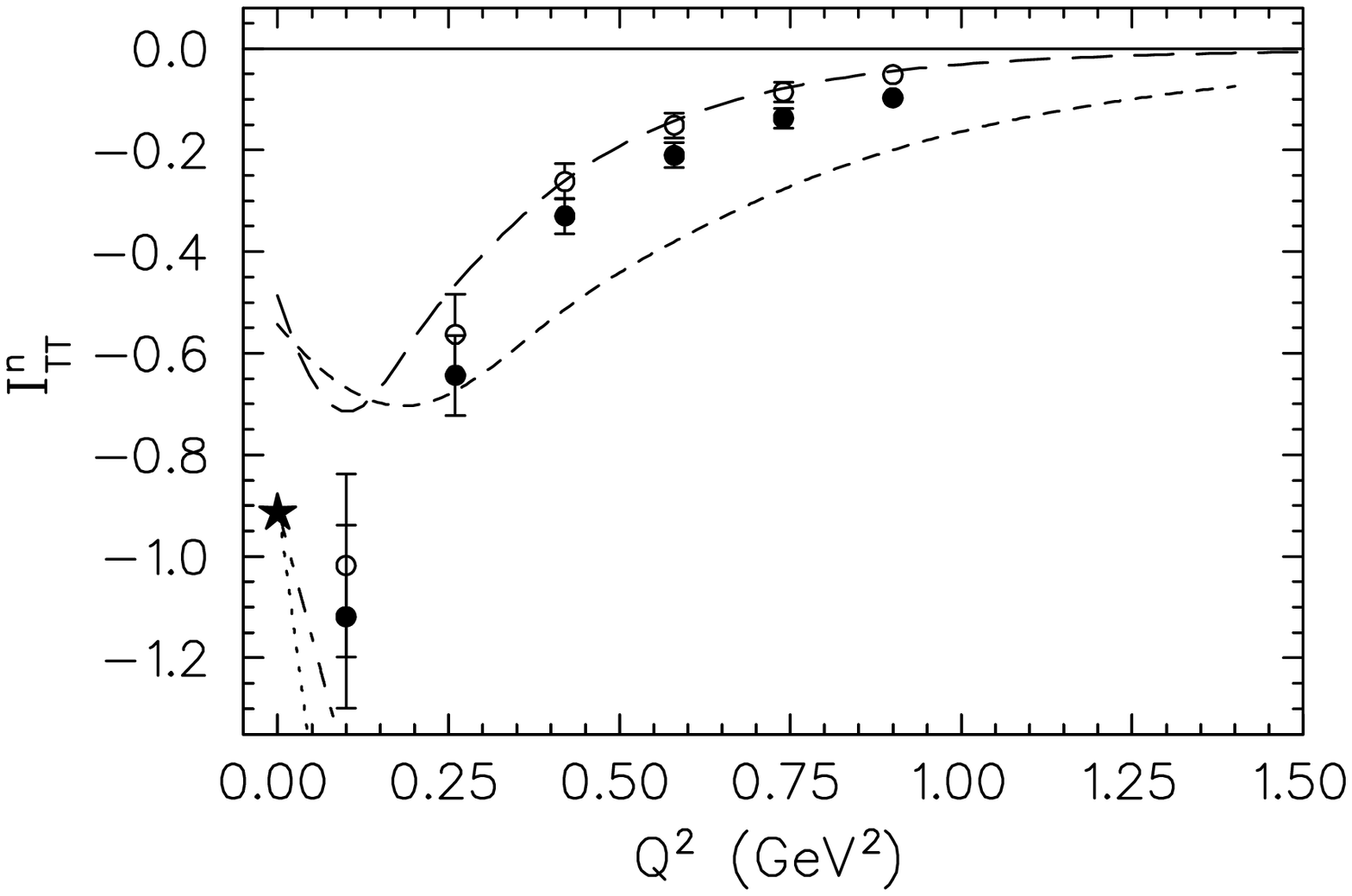,height=4.5cm}
\psfig{figure=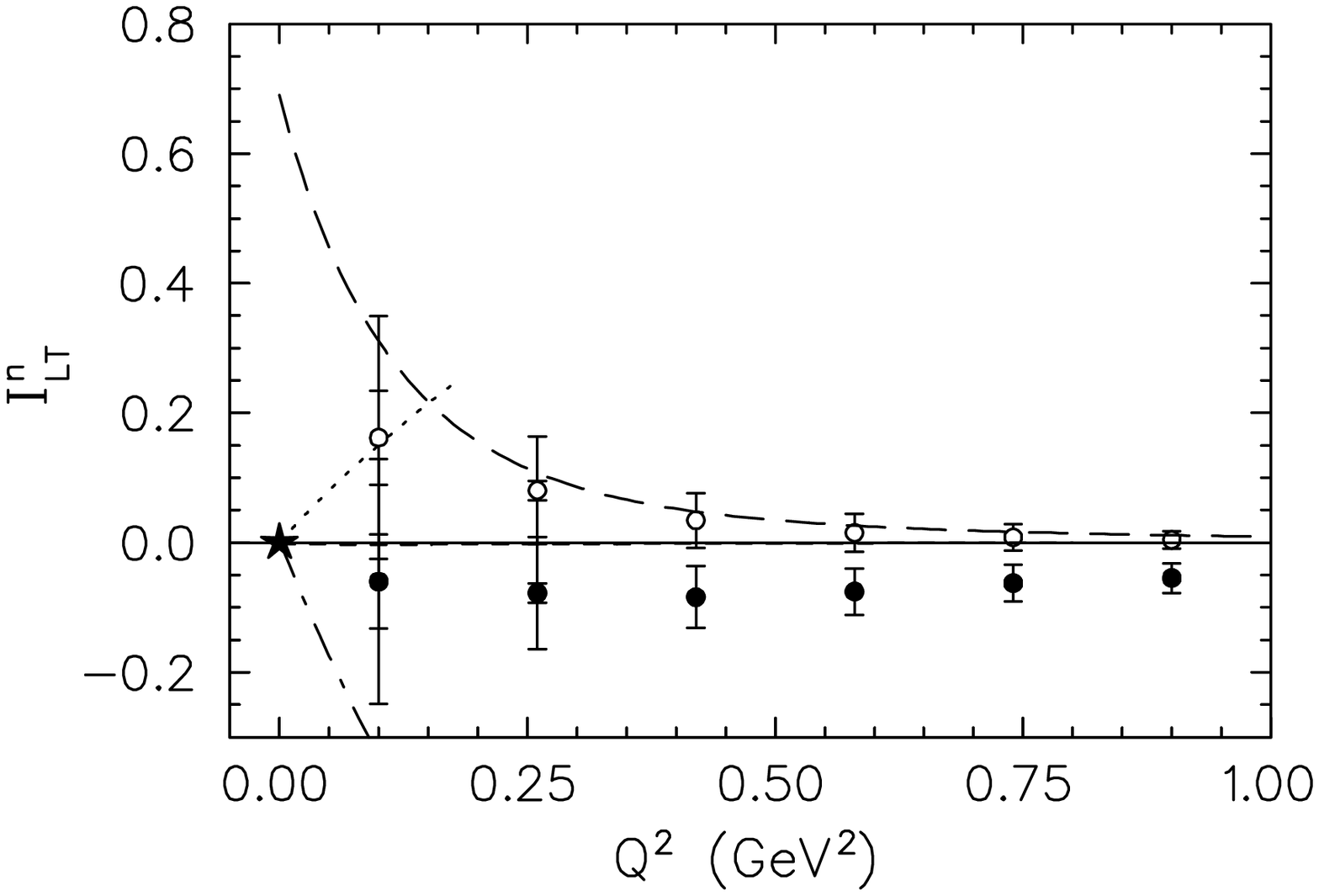,height=4.5cm}} \caption{The $Q^2$ dependence of the
neutron integrals $I_{TT}^n$ and $I_{LT}^n$ as defined by
Eqs.~(\ref{DDeq2.3.30}) and (\ref{eq:i3int}). The open circles show the
resonance contribution ($W<2$~GeV) as measured by the E 94-010 Collaboration at
JLab~\cite{Ama02,Ama03}, the solid circles are obtained by including an
estimate for the DIS contribution. The short-dashed lines are the results of
the hypercentral constituent quark model~\cite{Gor04}. See Fig.~\ref{fig:i1p}
for further notation. }
\label{fig:ian}
\end{figure}

Figure~\ref{fig:ian} displays the integrals $I_{TT}^n$ and $I_{LT}^n$ as
derived from the $^3$He data of the JLab E94-010
Collaboration~\cite{Ama02,Ama03} and corrected for nuclear effects according to
the procedure of Ref.~\cite{Cio97}. The comparison of the data for $I_{TT}^n$
with the MAID results exposes the same problem as already discussed for real
photons: The neutron helicity difference in the resonance region is not
properly described by the one-pion contribution as constructed from the phase
shift analyses, particularly at small values of $Q^2$. However, the strong
curvature at $Q^2\approx 0.1$~GeV$^2$ and also the predictions for
$Q^2\gtrsim0.4$~GeV$^2$ agree with the data quite well. At this point we can
only speculate about the missing strength in MAID. The discrepancy seems to be
a long-range phenomenon that disappears rapidly with increasing resolution, and
it is therefore quite unlikely that DIS contributions are responsible. The
remaining candidates are final-state interactions leading to coherent pion
production on the $^3$He target, possible modifications of the multipole
expansion for the bound neutron, or a surprisingly large two-pion component in
the neutron case.

The integral $I_{LT}^n = I_1^n+I_2^n$ is shown in the right panel of
Fig.~\ref{fig:ian}. It can be easily constructed from the previous results.
This observable deserves particular attention, because it samples the
information from the little known longitudinal-transverse cross section. As
Eq.~(\ref{eq:i3int}) indicates, the convergence of $I_{LT}$ requires that
$\sigma_{LT}$ drop faster than $1/\nu$ at large $\nu$. Because the
longitudinal-transverse interference involves a helicity flip, this is very
likely to happen at sufficiently large $\nu$. However, there is little
experimental information on $\sigma_{LT}$ over the whole energy region, and
therefore the phenomenological description is on rather shaky ground. The zero
of $I_{LT}^n$ in the real photon limit is particularly interesting (see
Eq.~(\ref{I_LT_0})), because it requires a complete cancellation of resonance
and DIS contributions. The nice agreement between the new JLab
data~\cite{Ama03} and MAID in the resonance region $(W<2$~GeV) may be called
semi-quantitative, except for the limit $Q^2\rightarrow0$ where both
experimental and theoretical error bars increase. Furthermore, the contribution
of DIS is large and negative over the full $Q^2$ region, which brings the
integral much closer to the predicted value at $Q^2=0$. Concerning the ChPT
calculations~\cite{JiK00a,Ber02}, the zero value at the photon point is of
course taken from Eq.~(\ref{I_LT_0}), but the steep slope is a genuine
prediction. In Figure~\ref{fig:ian} we also show the predictions of the
hypercentral constituent quark model~\cite{Gor04}, which are dominated by the
contributions of the $\Delta$ resonance. In this model, the longitudinal
strength of the $\Delta$ excitation is very small, and therefore the result for
$I_{LT}^n$ is practically zero over the whole range of $Q^2$.

\subsection*{4.5\quad Generalized Polarizabilities and Higher Twists}

In the previous section we have discussed the delicate cancellation between negative and
positive contributions to the GDH-like integrals, and in particular the rapid change of
the integrals as functions of momentum transfer. In the case of the generalized
polarizabilities $\gamma_{TT}(Q^2)$ and $\delta_{LT}(Q^2)$, the integrands are weighted
by an additional factor $\nu^{-2}$ or $x^2$, which enhances the importance of the
threshold region relative to resonance excitation, and suppresses the contributions of
the DIS continuum above $W=2$~GeV.

\begin{figure}
\centerline{\psfig{figure=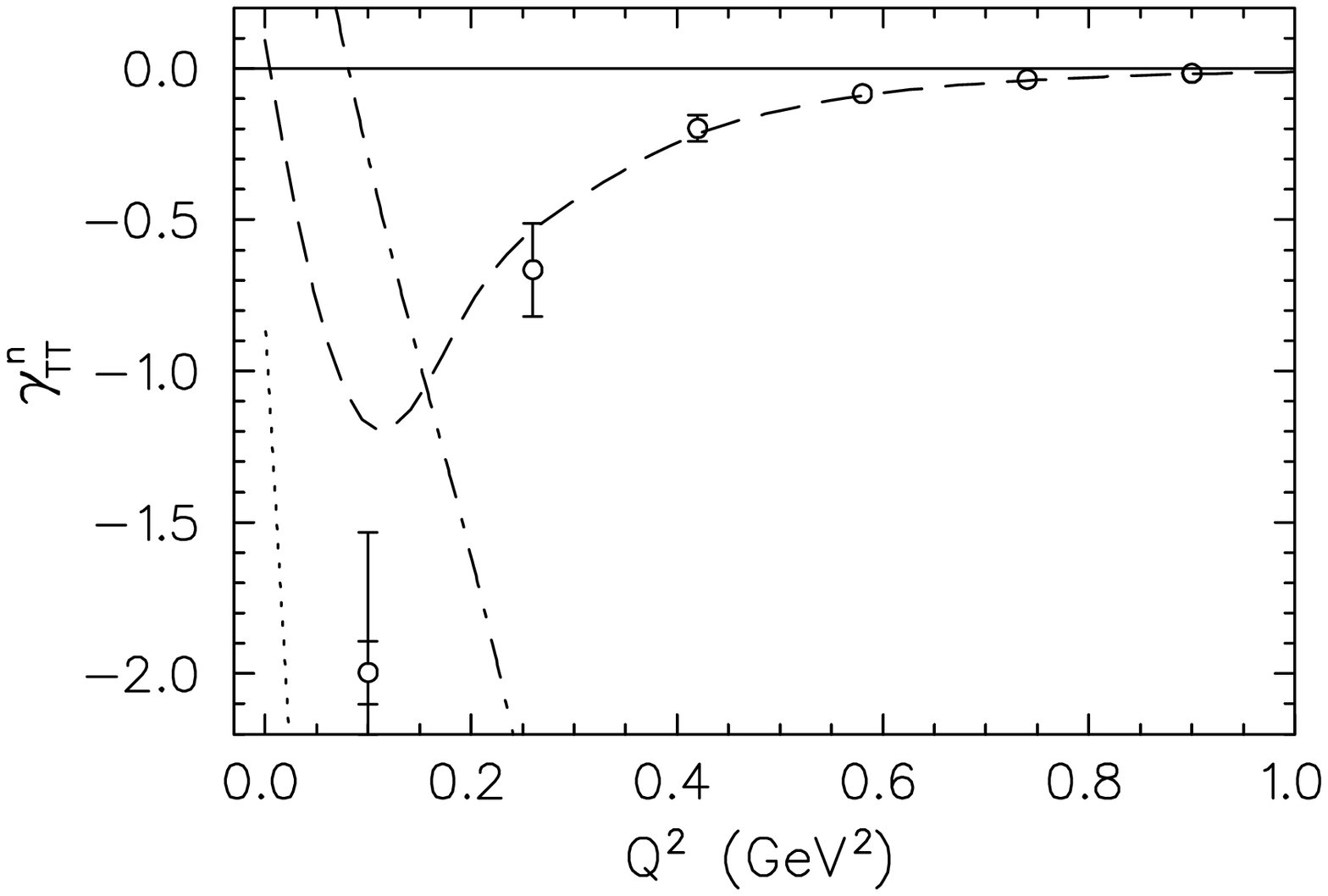,height=4.5cm}
\psfig{figure=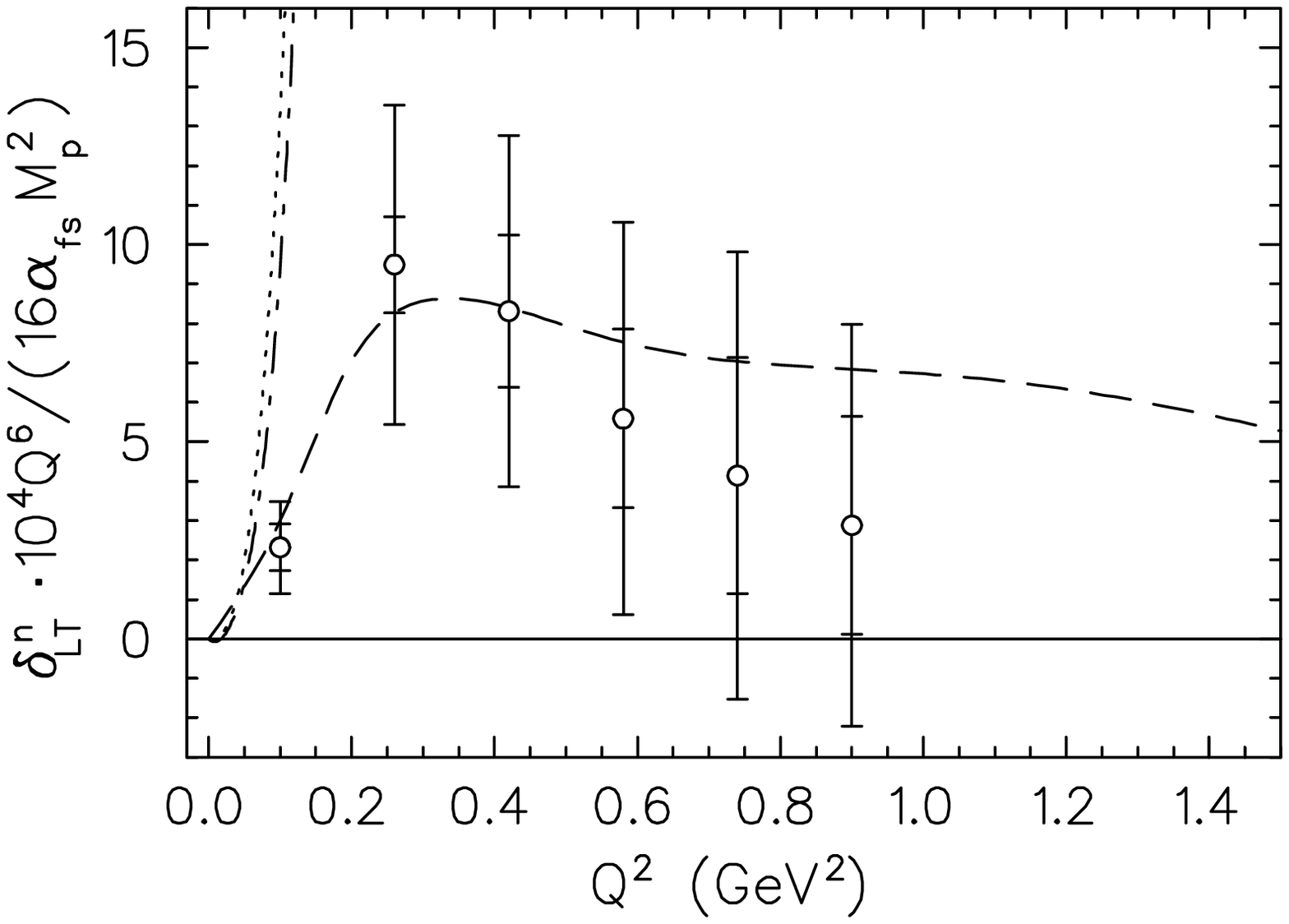,height=4.5cm}} \caption{ The $Q^2$ dependence of
the generalized neutron polarizabilities $\gamma_{TT}^n$ (left) and
$\delta_{LT}^n$ (right) as defined by Eqs.~(\ref{eq:gammao}) and
(\ref{eq:deltalt}). Note that $\delta_{LT}^n$ has been multiplied by a factor
of $10^4Q^6/(16\alpha_{fs}M_p^2)$ in order to compensate for its rapid decrease
with increasing $Q^2$. The open circles represent the preliminary data of the
JLab E 94-010 Collaboration~\cite{Cho04}. See Fig.~\ref{fig:i1p} for further
notation.
\label{fig:gammaTT} }
\end{figure}

Figure~\ref{fig:gammaTT} shows the FSP $\gamma_{TT}^n$ and the
longitudinal-transverse polarizability $\delta_{LT}^n$ (multiplied by a factor
$10^4Q^6/(16\alpha_{fs}M_p^2)$) as a function of $Q^2$. As in the case of
$I_{TT}^n$, the preliminary data for $\gamma_{TT}^n$~\cite{Cho04} show
considerably more strength at small $Q^2$ than predicted by MAID for the
one-pion channel. However, the agreement for $Q^2\gtrsim 0.4$~GeV$^2$ is again
quite satisfactory. In view of the additional weight factor towards the
low-energy region, this behavior is another indication that the ``neutron
problem'' should be related to low-energy and long-range phenomena.

A comparison with the predictions of ChPT shows once more that $\gamma_{TT}^n$
is a particularly sensitive observable. The phenomenological analysis yields
very small values of $\gamma_{TT}$ for both nucleons. The reason for this is
the large cancellation of s-wave pion production (multipole $E_{0^+}$) and
$\Delta$ resonance contributions (multipole $M_{1^+}$), as is immediately
apparant in Eq.~(\ref{sigma'TT}). Because of the additional weight factor
$\nu^{-2}$, this cancellation is considerably stronger than in the case of the
integral $I_{TT}$. It is therefore no big surprise that ChPT cannot describe
$\gamma_{TT}$ in a quantitative way. The ${\mathcal{O}}(p^3)$ and
${\mathcal{O}}(p^4)$ approximations of heavy baryon ChPT oscillate from
positive to negative values, whereas the FSP at $Q^2=0$ is essentially zero.
Also the newly developed Lorentz invariant version of ChPT misses the real
photon point, and the situation improves only slightly if $\Delta$ effects are
added by the resonance saturation method. This behavior changes in the case of
the longitudinal-transverse polarizability $\delta_{LT}$, which is dominated by
the s-wave term $S_{0+}^{\ast}E_{0+}$ in the multipole expansion of
Eq.~(\ref{sigmaLT}). The next term in that expansion, the multipole $1^+$, is
essentially saturated by the $\Delta(1232)$. However, because of the small
ratios $E_{1+}/M_{1+}$ and $S_{1+}/M_{1+}$ (see Section 4.3), this resonance
contribution is very much suppressed. Although the longitudinal strength of the
higher resonances is not well known, is is generally agreed to be small. As a
consequence $\delta_{LT}$ decreases rapidly as function of $Q^2$ without
showing any pronounced resonance structures. In other words, the
longitudinal-transverse polarization takes place in the outer regions of the
nucleon and is mostly due to the pion cloud. This notion is well supported by
the fact that the ChPT prediction for $\delta_{LT}$ is much better than the
prediction for $\gamma_{TT}$.

Up to this point we have concentrated on the polarizabilities, which are
defined by the low energy expansion of the VVCS amplitude
(Eqs.~(\ref{DDeq2.3.29}) and (\ref{eq:s1lex})) and expressed by moments of the
spin structure functions (Eqs.~(\ref{eq:gammao}) and (\ref{eq:deltalt})). We
now turn to the OPE and ask how the coefficients of this expansion evolve for
decreasing $Q^2$. The OPE yields the following relations~\cite{JiU94}:
\beqn \label{QM1} \int_0^1dx\ g_1(x,Q^2) & = & \tilde{\Gamma}_1 +
(a_2+4d_2+4f_2)\,\frac{M^2}{9Q^2}
 + {\mathcal{O}}\,\left(\frac{M^4}{Q^4}\right)\ , \\
\label{QM2}
\int_0^1dx\ g_2(x,Q^2) & = & {\mathcal{O}}\,\left(\frac{M^4}{Q^4}\right)\ , \\
\label{QM3} \int_0^1dx\ x^2g_1(x,Q^2) & = & \frac{1}{2} a_2 +
{\mathcal{O}}\,\left(\frac{M^2}{Q^2}\right)\ ,
\\
\label{QM4} \int_0^1dx\ x^2g_2(x,Q^2) & = & -\frac{1}{3} a_2 + \frac{1}{3} d_2 +
{\mathcal{O}}\,\left(\frac{M^2}{Q^2}\right)\ , \eeqn
with only a logarithmic $Q^2$ dependence of $\tilde{\Gamma_1},\ a_2,\ d_2$, and
$f_2$. In the usual nomenclature of the OPE, the leading term
$\tilde{\Gamma}_1$ is twist 2, $a_2$ is a target mass correction and also of
twist 2; and $d_2$ and $f_2$ are matrix elements of twist-3 and twist-4 quark
gluon operators, respectively. Combining Eqs.~(\ref{QM3}) and (\ref{QM4}) we
can define the function
\be
\label{d2Q2}
d_2(Q^2) = \int_0^1dx\ x^2\,\left(3g_2(x,Q^2) + 2g_1(x,Q^2)\right)\ , \ee
which approaches the twist-3 term $d_2$ of the OPE in the limit of
$Q^2\rightarrow\infty$. By use of the Wandzura-Wilczek sum rule,
Eq.~(\ref{d2Q2}) can be cast in the form
\be d_2(Q^2) = 3\int_0^1 dx\ x^2 \left(g_2(x,Q^2) -
g_2^{\rm{WW}}(x,Q^2)\right)\ , \ee
where $g_2^{\rm{WW}}$ is the twist-2 part of $g_2$ as obtained from Eq.~(\ref{eq:ww}).

Several model estimates as well as lattice QCD calculations have been performed
for the twist-3 matrix element $d_2$. In particular, an estimate in the
instanton vacuum approach~\cite{Lee02}, where $d_2$ is suppressed because of
the diluteness of the instanton medium, predicts that $d_2$ is of order
$10^{-3}$. Also a lattice calculation~\cite{Gock01} and the recent experimental
results from SLAC~\cite{E155X} support such a small value. Combining
Eqs.~(\ref{DDeq2.3.30},\ref{eq:deltalt},\ref{eq:I1}) with Eq.~(\ref{d2Q2}), we
find that the inelastic contribution to $d_2$ can be expressed as~\cite{Kao04}:
\begin{eqnarray}
d_2^{\rm{inel}}(Q^2) \,=\, {{Q^4} \over {8 M^4}} \left\{ I_1(Q^2) - I_{TT}(Q^2) +
\frac{M^2Q^2}{\alpha_{\mbox{\sc{em}}}} \, \delta_{LT}(Q^2) \right\} \, . \label{eq:d2b}
\end{eqnarray}
Because $I_{TT}(0) = I_1(0) = -\kappa^2/4$, the RHS of this equation is
determined by the slopes of the generalized GDH integrals at the real photon
point, and therefore $d_2(Q^2)$ can also be predicted by ChPT, at least for
sufficiently small $Q^2$.

\begin{figure}
\centerline{\psfig{figure=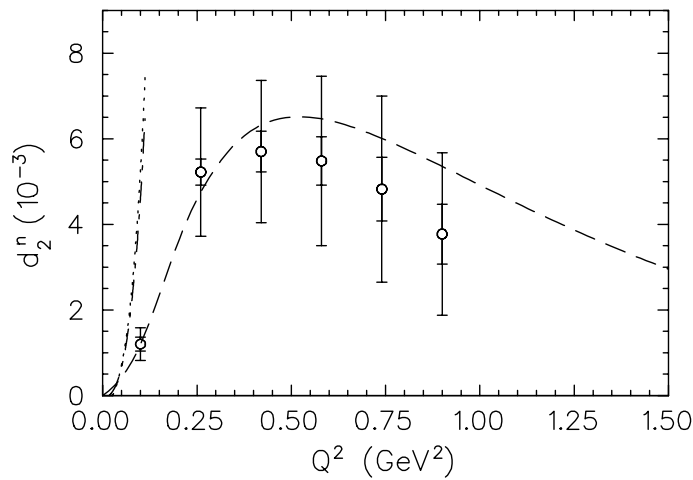,width=9cm}} \caption{ The inelastic
contribution to the neutron moment $d_2^n$ as defined in Eq.~(\ref{eq:d2b}).
The data are from JLab E94-010 (open circles, Ref.~\cite{Ama03}). See
Fig.~\ref{fig:i1p} for further notation. \label{d2n} }
\end{figure}

Figure~\ref{d2n} compares the recent data of the JLab E94-010
Collaboration~\cite{Ama03} for $d_2^n(Q^2)$ to the predictions of MAID and
heavy baryon ChPT~\cite{Kao04}. The observed value of order $10^{-3}$ is in
agreement with theory. By a similar analysis one can try to derive a value for
the matrix element $f_2$. Concerning the physics of these terms, it has been
suggested that the matrix elements $d_2$ and $f_2$ describe the response of the
color electric and magnetic fields to the polarization of the nucleon, which
can be expressed in terms of gluon field polarizabilities~\cite{FJ01},
\be d_2  =  \frac{1}{4}\,(\chi_E + 2\chi_B)\ , \quad f_2  =  (\chi_E - \chi_B) \ .
\label{polglue} \ee
Therefore an extraction of both $d_2$ and $f_2$ can yield interesting new nucleon
structure information. Such a phenomenological extraction was performed for the first
time in Ref.~\cite{JiMel97} on the basis of the SLAC data~\cite{Abe96}. A more recent
phenomenological analysis~\cite{Kao04} based on all the available proton data has
extracted a twist-4 matrix element $f_2\gg d_2$. From this relation and
Eq.~(\ref{polglue}) we obtain $\chi_E \approx \frac{2}{3}\,f_2$ and $\chi_B \approx  -
\frac{1}{3} \,f_2$. In particular, the positive value of $f_2$ found in the case of the
proton leads to a negative value of $\chi_B$, i.e., color diamagnetism.

\section*{5. \quad SUMMARY}

The GDH sum rule is one of several dispersive sum rules that connect the real
and virtual Compton scattering amplitudes to inclusive photo- and
electroproduction cross sections. Because these sum rules are based on such
universal principles as causality, unitarity, and gauge invariance, they
provide a unique testing ground for our understanding of the pertinent degrees
of freedom.

Recent experiments at MAMI and ELSA have shown a saturation of the GDH sum rule
for the proton at photon energies of $\sim3$~GeV, at least within the
experimental errors. The planned SLAC experiment will measure the helicity
cross sections up to 40~GeV, which should provide good leverage to test the
Regge predictions and the convergence of the integral.

What about the neutron sum rule? The analysis of the recently taken deuteron
data will help to answer this question. However, the problem of dividing the
nuclear response into the contributions of protons and neutrons will persist
for some time. In order to resolve this issue firmly, it will be helpful and
maybe even necessary to detect the hadronic final states. Altogether it would
be an awkward situation if the sum rule were to be fulfilled for the proton but
violated for the neutron.

The generalization of Compton scattering to virtual photons opens a wide field.
Deep inelastic lepton scattering at CERN, HERMES, and SLAC has yielded
invaluable information on the spin structure functions and their moments since
the beginning of the 1980s. In particular, the Bjorken sum rule, a strict
prediction of QCD, was found to agree with the data. However, a complete
picture of the spin dynamics in the nucleon requires the measurement of these
moments over the full range of momentum transfer, between the limits of
coherent and incoherent scattering. The Hall A and B Collaborations at JLab
have recently provided a rich body of high-quality data at small to
intermediate energy transfers, and more data will be forthcoming in the next
few years. In this review, we had to treat these new developments rather
cursorily. Let us therefore conclude by pointing out some of the highlights in
the field of virtual photons:

\begin{itemize}
\item The BC sum rule, derived on the basis of superconvergence, expresses
$I_2(Q^2)$, the first moment of the second spin structure function, in terms of
the ground state form factors $G_E(Q^2)$ and $G_M(Q^2)$. The data of JLab
E94-010 support this sum rule within the experimental and model errors. In
contrast to the GDH and Bjorken sum rules, this relation should be valid at
arbitrary momentum transfer $Q^2$.

\item Data for $I_1(Q^2)$ will be soon available for both proton and neutron
down to very small values of $Q^2$. As a result one will be able to construct
the isovector combination $I_1^p-I_1^n$ over a large range of momentum
transfer. This may open the possibility to bridge the gap between low-energy
(ChPT) and high-energy (perturbative QCD) calculations.

\item Preliminary data of JLab E94-010 exist for the two generalized spin
polarizabilitis $\gamma_{TT}^n$ and $\delta_{LT}^n$. Being functions of the
virtuality $Q^2$, these oberservables yield information on the spatial
distribution of the polarization densities of charge and magnetization. In the
case of the generalized FSP $\gamma_{TT}$, we observe a delicate cancellation
of pion s-wave production and $\Delta$ resonance excitation at the real photon
point. Because of the stronger form factor of the pion cloud contribution, the
cancellation becomes less effective at finite $Q^2$, and thus $\gamma_{TT}$
decreases towards a sharp minimum near $Q^2=0.1$~GeV$^2$. The
longitudinal-transverse polarizability $\delta_{LT}$, on the other hand, is
strongly dominated by the pion cloud and therefore drops rapidly with
increasing $Q^2$.

\item The new precision data in the resonance region allow us to determine the
higher twists of perturbative QCD by extrapolations of appropriate observables
from intermediate to large $Q^2$. This is particularly interesting in the case
of the twist-3 matrix element $d_2$, which measures the deviation from the
Wandzura-Wilczek-sum rule. Because such higher twists express the correlations
between the quarks due to gluon exchange, they may be interpreted as the
gluonic contributions to the polarizability of the nucleon.

\end{itemize}

In summary, considerable progress has been made in our qualitative
understanding of the nucleon's spin structure. The developments highlighted
here cover a broad range from the coherent to the incoherent response of the
nucleon. As the virtuality of the photon increases, this strongly correlated
many-body system is seen through a microscope with better and better
resolution, and the pertinent degrees of freedom change from Goldstone bosons
and collective resonances to those of the primary constituents, the quarks and
gluons. Over the past years, we have also witnessed a steady improvement of ab
initio calculations in the frameworks of ChPT, perturbative QCD and lattice
QCD. The wealth of the new data has been both informative and challenging for
theory, and we look forward to further experimental and theoretical advances in
our quest to understand the nucleon's spin structure in the realm of
non-perturbative QCD.

\section*{ACKNOWLEDGMENTS}
We are grateful to J.~Ahrens, H.J.~Arends, and P.~Pedroni for comments and
discussion. A special thank you goes to G.~Cates, J.P.~Chen, S.~Choi, and
Z.E.~Meziani for communicating preliminary results of the E94-010
Collaboration. We would also like to thank M.~Vanderhaeghen for carefully
reading the manuscript and contributing several useful and stimulating
suggestions. This work was supported by the Deutsche Forschungsgemeinschaft
(SFB 443).

\section*{LITERATURE CITED}

\end{document}